\documentclass[pdflatex,icol]{sn-jnl}

\usepackage{amssymb}
\usepackage{amsmath}

\usepackage{amsthm}
\usepackage{xurl}
\usepackage{natbib}
\usepackage{mathtools}
\usepackage[usenames,dvipsnames]{color}
\usepackage{bm}
\usepackage{bbm}
\usepackage{xcolor}
\usepackage{float}
\usepackage{commath}
\usepackage{booktabs}
\usepackage{tabularx}
\usepackage{graphicx}
\usepackage{multirow}
\usepackage{float}
\usepackage{makecell}
\usepackage{url}
\usepackage{enumitem}
\usepackage[linesnumbered]{algorithm2e} 
\RestyleAlgo{ruled} 

\renewcommand{\P}{\mathbb{P}}
\newcommand{\E}{\mathbb{E}}
\newcommand{\D}{\bm{\Delta}_\alpha }
\newcommand{\DD}{\bm{\Delta}_{\alpha2}}
\newcommand{\V}{\mathbb{V}}
\newcommand{\trace}{\text{trace}}
\newcommand{\cov}{\text{Cov}}

\newcommand{\argmin}{\operatornamewithlimits{argmin}}

\newcommand{\A}{\mathcal{A}}
\newcommand{\diag}{\text{diag}}

\theoremstyle{thmstyleone}%
\theoremstyle{thmstyletwo}%

\theoremstyle{thmstylethree}%

\begin{document}

\title[Robust Best Subset Selection via Fast Approximate MM-Estimation]{Robust Best Subset Selection via Fast Approximate MM-Estimation}


\author*[1]{\fnm{Martin} \sur{Huang}}\email{martin.huang@sydney.edu.au}

\author[1,2]{\fnm{Samuel} \sur{Muller}}

\author[1]{\fnm{Garth} \sur{Tarr}}

\affil[1]{\orgdiv{School of Mathematics and Statistics}, \orgname{The University of Sydney}, \orgaddress{\city{Sydney}, \postcode{2006}, \state{New South Wales}, \country{Australia}}}

\affil[2]{\orgdiv{Faculty of Science and Engineering}, \orgname{Macquarie University}, \orgaddress{\city{Sydney}, \postcode{2109}, \state{New South Wales}, \country{Australia}}}


\abstract{Best subset selection procedures typically rely on a squared error loss, where a small number of outlying observations may distort the entire solution path. Replacing this loss with a robust alternative, such as the MM-estimator robust model selection criterion, is computationally prohibitive, as it requires an iterative fit across near-exhaustive candidate model spaces. To preserve the robust properties of the MM-estimator while avoiding its repeated iterative computation, we introduce FAMM, a Fast Approximate MM-Estimator based on a weighted least squares fit with weights obtained from a full data MM-estimator. Although the resulting estimator is no longer an MM-estimator, we prove that model selection consistency is retained when aligning the weights of the selection criterion with those of the estimator. We embed FAMM within COMBSS, whose continuous relaxation of the discrete best subset search enables scalable best subset selection for large numbers of predictor variables. Consequently, robust model selection and best subset selection with large and contaminated datasets are made feasible.}


\maketitle
\section{Introduction}\label{chap:introduction}
Best subset selection (BSS) addresses the variable-selection problem in regression by, for each subset size $k$, choosing the $k$ predictors that yield the best predictive model for a response. This results in a best subset solution path, the collection of $p$ candidate models obtained by solving the BSS problem for each subset size $k=1,\ldots,p$, from which a final model can then be chosen. Since searching all subsets is computationally expensive, BSS becomes intractable beyond a small number of predictor variables $p$, motivating continuous relaxations of the discrete search. COMBSS is one such method, adjusting the binary vector space $\{0,1\}^p$ of exact BSS into a hyper-cube $[0,1]^p$, enabling standard continuous optimisation methods, such as gradient descent \citep{moka_combss_2024}. In both cases, determining the solution path and final model is not an easy task. In many applications, this choice is guided by a balance between in-sample fit, model complexity, and predictive performance. While COMBSS and BSS are built on squared error losses, such as the mean squared error (MSE), other criteria in general model selection include Akaike's Information Criterion (AIC), the Bayesian Information Criterion (BIC), Shao's Criterion, and Mallows' $C_p$ \citep{akaike_information_1998, mallows_comments_1973, schwarz_estimating_1978, shao_bootstrap_1996}.

In the presence of outliers, this reliance on squared error loss is well known to produce poor model rankings, motivating substantial literature on robust model selection, where classical estimators are replaced by robust extensions \citep{guney_robust_2021,ronchetti_robust_1985, sommer_variables_1996,Thas_2026}. The same idea appears in the robust BSS literature, where the loss underlying the subset search is also replaced \citep{christidis_robust_2026,thompson_robust_2022}. Although this extension is conceptually straightforward, it presents substantial computational challenges.

Many criteria are approximated through the bootstrap, where predictive measurements are assessed by repeated train/test splits or cross-validation \citep{kepplinger_information_2026, muller_outlier_2005,shao_bootstrap_1996, werner_trimming_2023}. Classical resampling-based procedures, however, can misrepresent the proportion of outliers, since some samples may contain too many or too few. To address this issue, \citet{muller_outlier_2005} proposed a stratified bootstrap, in which observations are grouped according to robust regression residuals, and resampled within strata, such that the proportion of outliers can be controlled across bootstrap samples. This stratification stabilises the resampling distribution, and provides a more representative bootstrap sample for robust model selection.

For every candidate model per bootstrapped sample, a robust criterion typically requires fitting iterative estimators, such as an M- or MM-estimator \citep{koller_nonsingular_2023, muller_outlier_2005, ronchetti_robust_1997, smucler_robust_2017}. Since these estimators do not have closed-form solutions, the computational burden can be substantial. This is especially pronounced when considering that BSS performs model selection on upwards of $2^p$ models, and COMBSS makes solution paths tractable for large $p$. Naively substituting an MM-estimator-based criterion in COMBSS would counteract the computational advantage that makes COMBSS attractive to begin with.

To retain the robustness of the MM-estimator without forfeiting COMBSS's computational scalability, we introduce Fast Approximate MM-Estimation (FAMM), a weighted least squares (WLS) approximation to the MM-estimator that can be embedded directly within COMBSS's candidate search.

FAMM computes a single MM-estimator on the full dataset, and uses the constructed weights to fit the much faster closed-form WLS estimator on each bootstrap sample. As the MM-estimator is now approximated through WLS, FAMM is especially fast, does not sacrifice performance, and can be easily integrated into COMBSS. In small to moderate dimensional settings, FAMM also makes exhaustive searches feasible. For example, in Section \ref{sec:NBA}, FAMM completes an exhaustive search over $2^{12} = 4096$ models in approximately four seconds. Without the approximation, running the same search with the MM-estimator takes about 90 minutes, over a 1000-fold slowdown, while producing an identical result. Our contribution, however, is not solely computational. Since the weights are held fixed across bootstrap samples, the resulting estimator is no longer an MM-estimator, meaning the consistency results from \citet{muller_outlier_2005} do not transfer automatically. In Section \ref{sec:theory}, we show that by aligning the model selection criterion weights with the estimator's robustness weights, model selection consistency is retained.

\citet{salibian-barrera_bootstrapping_2002}, and later \citet{salibian-barrera_robust_2008} introduced the Fast and Robust Bootstrap (FRB), which closely aligns with the goals of FAMM. Instead of refitting the MM-estimator on each bootstrap, it calculates a reweighted representation, whilst computing a linear correction to account for the loss of variability in the bootstrap estimator. This correction is a key factor in constructing bootstrap confidence intervals and point estimation. When focusing on model selection rather than confidence interval estimation or inference, our empirical simulations in Section \ref{chap:simulations} do not suggest that this correction is essential. Including the correction steps also increases computational cost while yielding only marginal improvements in model selection performance, in comparison to the slow gold standard of MM-estimation. 

The remainder of the paper is structured as follows. Section \ref{chap:preliminaries} formalises the notation for BSS and COMBSS, and introduces the statistical background for robust model selection used throughout the paper. Section \ref{chap:rCOMBSS} describes the proposed FAMM and Robust COMBSS methodology, along with the conditions required for robust model selection consistency. Section \ref{chap:simulations} presents the results from an extensive simulation study, assessing the empirical appropriateness of FAMM as an approximation to the MM-estimator, and benchmarking its model selection performance and computational cost. Finally, Section \ref{chap:conclusion} concludes the paper.

\section{Preliminaries}\label{chap:preliminaries}
In this paper, we focus on linear regression models. Let $n$ be the number of independent observations, let $\bm{y} = (y_1, \dots, y_n)^{\top}$ be the response vector, and let $\bm{X} \in \mathbb{R}^{n\times p}$ be the design matrix with columns indexed by $\{1,\dots,p\}$. For any subset of indices $\alpha \subseteq \{1,\dots,p\}$, we write $\bm{X}_\alpha \in \mathbb{R}^{n\times p_\alpha}$ for the corresponding submatrix of $\bm{X}$, where $p_\alpha$ is the number of variables in $\alpha$, and we write $\bm{x}_{\alpha i}^{\top}$ for the $i$th row of $\bm{X}_\alpha$. When $\alpha$ contains all indices, we refer to it as the full model and write $\alpha_f$. The true model is denoted as $\alpha_0$, and the set of all candidate models is denoted as $\mathcal{A}$.

\subsection{Best Subset Selection and COMBSS}

BSS aims to identify the best subset solution path by solving $$\min_{\bm{\beta} \in \mathbb{R}^p} \frac{1}{n} \lVert \bm{y}-\bm{X}_{\alpha_f}\bm{\beta}\rVert^2 _2,\ \text{s.t. } \lVert \bm{\beta}\rVert_0 = k,$$ for a given subset size $k$. For a binary vector $\bm{s} = (s_1, \dots , s_p) \in \{0,1\}^p$, the exact best subset selection problem is 
\begin{align}\label{eq:dcombss}
\min_{\bm{s} \in \{0,1\}^p} \frac{1}{n} \lVert \bm{y} - \bm{X}_{\bm{s}} \hat{\bm{\beta}}^{\text{OLS}}_{\bm{s}} \rVert^2 _2,\ \text{s.t. }|\bm{s}| = k,    
\end{align}

\noindent where $\bm{X}_{\bm{s}}$ is the subset of $\bm{X}_{\alpha_f}$ containing only the columns $j$ for which $s_j = 1$, and $\hat{\bm{\beta}}^{\text{OLS}}_{\bm{s}}$ is the corresponding OLS estimator. COMBSS, a continuous optimisation method for best subset selection, extends the best subset solution path by adjusting the binary vector space $\bm{s} \in \{0,1\}^p$ to a whole hyper-cube $[0,1]^p$ for each $\bm{t}\in [0,1]^p$ \citep{moka_combss_2024}. In turn, a well-defined continuous extension of the exact problem in (\ref{eq:dcombss}) can be expressed as 
\begin{align}
    \min_{\bm{t} \in [0,1]^p} \frac{1}{n}\lVert\bm{y} - \bm{X}_{\bm{t}} \tilde{\bm{\beta}}_{\bm{t}}\rVert^2 _2 + \lambda \sum^p _{j=1} t_j.
\end{align}
The new design matrix $\bm{X}_{\bm{t}}$ is obtained by multiplying the $j$th column of $\bm{X}_{\alpha_f}$ by $t_j$, for $j = 1,\dots,p$. The regularisation parameter $\lambda$ controls the sparsity level, and $\tilde{\bm{\beta}}_{\bm{t}}$ is the new coefficient estimator, as defined in \citet[Equation (7)]{moka_combss_2024}.

COMBSS applies a gradient descent algorithm to this objective function and stores the resulting optimisation path in $\bm{t}_{\text{path}}$ \citep[Algorithm 1]{moka_combss_2024}. Each element $\bm{t}$ of $\bm{t}_\text{path}$ produces at most one model, for each model size $k \in \{1,\dots,p-1\}$. Since the number of models of size $k$ in $\bm{t}_{\text{path}}$ is equal to the number of iterations $l$ in the optimisation algorithm, there are at most $l$ models per size $k$.

Extending the difficult discrete constrained BSS problem to an unconstrained continuous optimisation problem allows for smaller computational cost, enabling scalability for larger $p$. While COMBSS is inherently faster than BSS, they are both built on a squared error loss, such that a single outlying observation can distort the entire solution path. This motivates replacing the OLS estimator with a robust alternative, which we introduce next.

\subsection{MM-Estimation and the Weighted Least Squares Representation}

The robust estimator we adopt to make COMBSS robust is the MM-estimator, which combines a high breakdown point with high efficiency, and whose weighted least squares representation, introduced later, is the key property we later utilise to construct a fast robust loss for BSS \citep{yohai_high_1987}.

For a given model $\alpha$, the MM-estimator satisfies
\[
\frac{1}{n} \sum_{i=1}^n \psi \left(\frac{y_i - \bm{x}_{\alpha i}^\top \hat{\bm{\beta}}_{\alpha}}{\hat{\sigma}_{\alpha}} \right) \bm{x}_{\alpha i} = \bm{0},
\]
where $\psi$ is the derivative of the loss function $\rho_1$ and $\hat{\sigma}_{\alpha}$ is an S-estimate of scale constructed with model $\alpha$ \citep{rousseeuw_robust_1984,yohai_high_1987}. When $\psi$ is a smooth $\psi$-function as in \citet{yohai_high_1987}, the MM-estimator can be represented as the WLS normal equation, $\sum_{i=1}^n v_{\alpha i} \bm{x}_{\alpha i} (y_i - \bm{x}_{\alpha i}^\top \hat{\bm{\beta}}_{\alpha}) = \bm{0}$, where the weights satisfy $v_{\alpha i} = V(r_{\alpha i}/\hat{\sigma}_{\alpha})$ with
\begin{align*}
    V(r_{\alpha i}/\hat{\sigma}_{\alpha}) = \begin{cases}
    \frac{\psi(r_{\alpha i}/\hat{\sigma}_{\alpha})}{r_{\alpha i}/\hat{\sigma}_{\alpha}} & \text{if } r_{\alpha i}/\hat{\sigma}_{\alpha}\neq 0, \\
    \psi^\prime (0)& \text{if } r_{\alpha i}/\hat{\sigma}_{\alpha} = 0,
\end{cases}
\end{align*}
and residuals $r_{\alpha i} = y_i - \bm{x}_{\alpha i} ^\top \hat{\bm{\beta}}_{\alpha}$. The MM-estimator may be written in WLS form as
\begin{align}\label{eq:wls}
\hat{\bm{\beta}}_{\alpha} = (\bm{X}^\top_\alpha \bm{V}_\alpha \bm{X}_\alpha)^{-1} \bm{X}^\top_\alpha \bm{V}_\alpha \bm{y},
\end{align}
where $\bm{V}_\alpha$ is a diagonal matrix of robust weights with entries $\diag(\bm{V}_{\alpha}) = (v_{\alpha 1}, \dots, v_{\alpha n})$ \citep{maronna_robust_2006}. Since the weights depend on the residuals which themselves depend on $\hat{\bm{\beta}}_{\alpha}$, the estimator must be computed iteratively, typically by iteratively reweighted least squares (IRLS) \citep{yohai_high_1987}. The IRLS procedure produces both the estimate $\hat{\bm{\beta}}_{\alpha}$ and the converged weights $v_{\alpha 1}, \dots, v_{\alpha n}$ as outputs.

A central computational feature of our method, presented in Section~\ref{chap:rCOMBSS}, is that we compute these weights only once---for a single model $\alpha_f$---and reuse them across all candidate models $\alpha \in \mathcal{A}$. When the weights $\bm{V}_{\alpha_f}$ are held fixed and applied to a different model $\alpha$, (\ref{eq:wls}) becomes a fixed-weight WLS estimator rather than a true MM-estimator. Since the weight matrix is then common to every candidate model, we drop the subscript for simplicity, and write $\bm{V} := \bm{V}_{\alpha_f}$ and $\diag(\bm{V}) = (v_{ 1}, \dots, v_{ n})$ throughout the remainder of the paper, in which $\alpha$ enters only through the design matrix $\bm{X}_\alpha$. The same argument follows for $\sigma\coloneqq \sigma_{\alpha_f}$ and $\hat{\sigma}\coloneqq \hat{\sigma}_{\alpha_f}$. 

\subsection{Robust Model Selection Criterion}

There is no consensus methodology for determining an optimal model from a given set of candidate models. One researcher may prefer models that contain a small set of predictor variables, such that they are easily interpretable and relatively simple. Another may decide to forfeit interpretation to prioritise prediction accuracy. \citet{muller_outlier_2005} suggest two indicators for identifying a useful linear regression model: (a) the model should parsimoniously describe the relationship between the sample data $\bm{y}$ and $\bm{X}$ and (b) be able to predict independent new observations. Following this philosophy, \citet{muller_outlier_2005} proposed the model selection criterion:
\begin{align}\label{eq:MSC}
M(\alpha) = \frac{\sigma^2}{n}\left(\E \sum^n _{i=1} w_{i} \rho \left \{\frac{y_i - \bm{x}^\top _{\alpha i} \hat{\bm{\beta}}_{\alpha}}{\sigma}\right\} + \delta (n) p_{\alpha} + \E \left [ \sum^n _{i=1} w_{i} \rho \left \{ \frac{z_i - \bm{x}^\top _{\alpha i}\hat{\bm{\beta}}_{\alpha}}{\sigma}\right\} \Bigg| \bm{y}, \bm{X} \right] \right).
\end{align}

An optimal model with subset indices $\alpha$ out of a set of candidate models $\mathcal{A}$ is then selected as the one that minimises the criterion $M(\cdot)$, denoted by $\hat{\alpha} = \argmin_{\alpha \in \A} M(\alpha)$. The criterion in (\ref{eq:MSC}) includes a weight parameter $w_{i}$, typically chosen to be $1$, or chosen to match the weights given in a Mallows M-estimator \citep{muller_outlier_2005}. The parameter $\rho$ is a loss function, recommended to be a trimming function, or a hard rejection function, such as $\rho(z) = \min(z^2, b^2)$ with constant $b$. The parameter $\sigma$ is the scale of the error distribution of a full model, that is, when $\alpha$ includes all indices in $\bm{X}$. The latter expectation captures future prediction error with $\bm{z} = (z_1, \dots, z_n)^\top$ as a vector of future responses. Finally, the complexity parameter $\delta(n) p_\alpha$ penalises the number of predictors $p_\alpha$, and the sample size $n$ through a function $\delta(n)$, typically chosen as $2\log(n)$.  

The former expectation in (\ref{eq:MSC}) is a penalised in-sample term, and can be explicitly computed. The latter expectation is the conditional expected prediction loss, and is estimated through an $m$ out of $n$ stratified bootstrap. The scale parameter $\sigma$ is estimated as the median absolute deviation (MAD) of the residuals with the full model MM-estimator $r_{\alpha_f i} = y_i - \bm{x}^\top _{\alpha_f i }\hat{\bm{\beta}}_{\alpha_f}$ \citep{hampel_1974}.

A stratified bootstrap is used to reduce the occurrence of bootstrap samples containing too many or too few outliers. Out of $H$ strata, observations with large residuals are allocated to lower and upper tail strata, whilst the remaining observations are placed between the tails. Sampling $m/H$ observations with replacement from each stratum then preserves a more stable proportion of outliers across bootstrap samples. The bootstrap estimate equivalent of (\ref{eq:MSC}) is then written as 
\begin{align}\label{eq:eMSC}
M_n(\alpha) = \frac{\hat{\sigma}^2}{n}\left( \sum^n _{i=1} w_i\rho \left \{\frac{y_i - \bm{x}^\top _{\alpha i} \hat{\bm{\beta}}_{\alpha}}{\hat{\sigma}}\right\} + \delta (n) p_{\alpha} + \E_* \sum^n _{i=1}w_i \rho \left \{ \frac{y_i - \bm{x}^\top _{\alpha i}\hat{\bm{\beta}}^*_{\alpha}}{\hat{\sigma}}\right\}\right),
\end{align}
where $\E_*$ is the expectation with respect to the empirical bootstrap distribution, and $\hat{\bm{\beta}}^* _\alpha$ is the MM-estimator computed on the bootstrap sample. The model that minimises $M_n(\alpha)$ is then denoted $\hat{\alpha}_{m,n} = \argmin_{\alpha \in \mathcal{A}} M_n (\alpha)$.

\section{Robust Best Subset Selection via Fast Approximate MM-Estimation} \label{chap:rCOMBSS}
In this section, we introduce FAMM, a computationally efficient approximation to the MM-estimator which enables the practical use of Robust COMBSS. 

\subsection{Fast Approximate MM-Estimation}
Evaluating the robust model selection criterion in (\ref{eq:eMSC}) requires a bootstrap estimator to be fit for every candidate model and every bootstrap sample, which is computationally prohibitive as an MM-estimator is iteratively refit at each step. This is particularly profound across large candidate spaces generated by COMBSS. FAMM mitigates the expensive computational cost by computing a single MM-estimator on the full dataset, extracting the associated weights, and using the fixed weights to fit WLS estimators on each bootstrap sample. 

This approximation is motivated by the WLS representation of the MM-estimator in (\ref{eq:wls}); however, it is important to distinguish the MM-estimator's WLS representation from the estimator used by FAMM. For a given model $\alpha$, the MM-estimator can be written as a WLS estimator only when the weights are induced by the residuals from that same model. FAMM instead fixes the weights obtained from the full-model MM fit and reuses them for candidate models and bootstrap samples. Consequently, the FAMM estimator for a candidate model is a full-model reweighted WLS estimator, designed to approximate the behaviour of the bootstrap MM-estimator selection criterion, rather than the exact MM-estimator under that candidate model.

By bootstrapping the weights alongside the data, influential observations in the bootstrap samples are subsequently down-weighted, and each bootstrap estimator can now be obtained through a much faster WLS computation. To control the proportion of outliers across bootstrap samples, FAMM is implemented within the stratified bootstrap scheme described in Section \ref{chap:preliminaries}. The resulting bootstrap estimators are then used to evaluate the robust model selection criterion in (\ref{eq:eMSC}).

In Algorithm \ref{alg:FAMM}, it is shown that the initial full MM-estimator contains the majority of the computational cost in FAMM. Subsequent bootstrap computations are extremely fast, due to the closed-form solution of the WLS.

As the weight matrix is fixed during the bootstrap procedure, FAMM under-represents the variance among the bootstrapped estimators, leading to anti-conservative confidence intervals. \citet{salibian-barrera_bootstrapping_2002} solve this issue by providing a linear correction to the bootstrap estimator in their FRB procedure. However, in our simulated analysis in Section \ref{chap:simulations}, this correction does not affect the performance in robust model selection. We find the simpler FAMM provides a balanced compromise between computational speed and model selection performance. Even if a practitioner wanted to construct confidence intervals or make predictions, they can do post-selection inference on the FAMM-optimal model.

\let\oldnl\nl
\SetNlSty{texttt}{(}{)}
\SetAlgoVlined
\begin{algorithm}
\KwIn{$\bm{X}_{\alpha_f}, \bm{y}, B, \mathcal{A}, H,m,\rho,\delta(n),\bm{w}$. \textbf{Optional}: $\bm{V},\bm{r}_{\alpha_f}$.}
\tcp{Initial MM-estimation Step}
\If{$\bm{V}$ \textnormal{and} $\bm{r}_{\alpha_f}$ \textnormal{not provided}}{
$\hat{\bm{\beta}} _{\alpha_f} \gets \text{Compute MM-estimation on } (\bm{X}_{\alpha_f}, \bm{y})$\;
$v_i \gets \text{Extract weights from IRLS for observations }i \in\{ 1,\dots,n\}$\;
$\bm{V} \gets \diag(v_1,\dots, v_n)$\;
$\bm{r}_{\alpha_f} \gets \bm{y} - \bm{X}_{\alpha_f} \hat{\bm{\beta}}_{\alpha_f}$\;
}

\For{$\alpha \in \mathcal{A}$}{
$\hat{\bm{\beta}} _{\alpha} \gets (\bm{X}^{\top} _\alpha \bm{V} \bm{X}_{\alpha} )^{-1} \bm{X}^{\top}_{\alpha} \bm{V} \bm{y}$\;
}
\vspace{1em}
\tcp{Stratification Step}
$\hat{\sigma} \gets$ MAD$(\bm{r}_{\alpha_f})$\;
Obtain order statistics of $\bm{r}_{\alpha_f}$: $r_{n, 1} \leq \cdots \leq r_{n, n}$\;
With the number of strata $H$, fix the percentiles $q_0 = 0,\ q_1, \dots, q_{H-1},\ q_H = 1$, where $q_h - q_{h-1} = 1/H$\;
\For{$h = 1$ \KwTo $H$}{
\For{$i = 1$ \KwTo $n$}{
    \If{$r _{\alpha_f i}\in [r _{n,[nq_{h-1}]+1} , r _{ n, [nq_h] + 1}]$}{
    $\text{Input }(y_i, \bm{x}_{i}, v_i) \rightarrow \mathcal{Q}_h$\;
    }
}
$m_h \gets m/H$\;
}

\vspace{1em}
\tcp{Bootstrapping Step}
\For{$\alpha \in \mathcal{A}$}{
    $L_{\text{in}} (\alpha) \gets \sum^n _{i=1} w_i\rho\left\{\frac{y_i - \bm{x}_{\alpha i}^\top \hat{\bm{\beta}}_{\alpha}}{\hat{\sigma}}\right\}$\;
    $L_{\text{boot}}(\alpha) \gets 0$
}

\For{$j=1$ \KwTo $B$}{
    $\bm{X}^* \gets \emptyset;\ \bm{y}^* \gets \emptyset;\ \bm{V}^* \gets \emptyset$\;
    \For{$h=1$ \KwTo $H$}{
        $(\bm{X}^* _h, \bm{y}^* _h,\bm{V}^* _{h}) \gets$ Sample $m_h$ rows from $\mathcal{Q}_h$ with replacement\;
        Append $(\bm{X}^* _h, \bm{y}^* _h,\bm{V}^* _{h})$ to $(\bm{X}^*, \bm{y}^*, \bm{V}^*)$\;
    }
    \For{$\alpha \in \mathcal{A}$}{
        $\hat{\bm{\beta}}^{*} _{\alpha} \gets (\bm{X}^{*\top} _\alpha \bm{V}^*  \bm{X}^*_{\alpha} )^{-1} \bm{X}^{*\top}_{\alpha} \bm{V}^*\bm{y}^*$\;
        $L_{\text{boot}}(\alpha) \gets L_{\text{boot}}(\alpha) + \sum^n _{i=1}w_i\rho \left\{\frac{y_i - \bm{x}_{\alpha i}^\top \hat{\bm{\beta}}^* _{\alpha}}{\hat{\sigma}}\right\}$\;
    }
}

\For{$\alpha \in \mathcal{A}$}{
    $M_n(\alpha) \gets \frac{\hat{\sigma}^2}{n}(L_{\text{in}}(\alpha) + \delta(n)p_\alpha + L_{\text{boot}}(\alpha)/B)$
}
$\hat{\alpha}_{m,n} \gets \argmin_{\alpha \in\A}M_n (\alpha)$\;
\KwRet{$\hat{\alpha}_{m,n}$}
\caption{FAMM}\label{alg:FAMM}
\end{algorithm}

\subsection{Consistency}\label{sec:theory}
As FAMM holds the weights of an MM-estimator fixed across candidate models and bootstrap samples, it is no longer a true MM-estimator. Consequently, the model selection consistency established in \citet[Theorem 1.]{muller_outlier_2005} does not automatically transfer. This section demonstrates that FAMM is model selection consistent when the weights of the selection criterion in (\ref{eq:eMSC}) match the robust MM-estimator weights, that is, $w_i = v_i$. 

Model selection consistency in the sense that $\lim_{n\rightarrow \infty} \P \{\hat{\alpha}_{m,n} = \alpha _0 \} = 1$ can be obtained when six conditions are satisfied, labelled (A) -- (F) in \citet[Section 2.]{muller_outlier_2005}.

 Only Condition (C) directly restricts the type of bootstrap estimator. Therefore, it is sufficient to verify Condition (C) holds, by utilising the estimator assumptions in Conditions (A) and (B). We restate these three conditions below for the particular case of FAMM.

\begin{enumerate}[label = (\Alph*)]
    \item The $p_\alpha \times p_\alpha$ matrix $$\bm{\Gamma}_\alpha = \lim_{n\rightarrow \infty} n^{-1} \sum^n _{i=1} w_{i} \bm{x}_{\alpha i} \bm{x}_{\alpha i}^\top$$ is of full rank and $n^{-1} \sum^n _{i=1} |w_{i}|^2 |\bm{x}_i|^4 <\infty$.
    \item For all models $\alpha \in \A$, the MM-estimator $\hat{\bm{\beta}}_\alpha - \bm{\beta}_\alpha = O_p(n^{-1/2})$ and $\hat{\sigma} - \sigma = O_p(n^{-1/2})$ with $\sigma >0$. For all correct models $\alpha \in \A_c$,
    $$n\V(\hat{\bm{\beta}}_\alpha) = \phi \bm{\Delta}_\alpha ^{-1} \bm{\Delta}_{\alpha2}\bm{\Delta}^{-1}_\alpha + o_p(1)$$
    for a scalar $\phi$, and \[\D = \lim_{n\rightarrow \infty} n^{-1} \sum^n _{i=1} v_{\alpha i} \bm{x}_{\alpha i}\bm{x}_{\alpha i}^\top,\ \ \DD = \lim_{n\rightarrow \infty} n^{-1} \sum^n _{i=1} v_{\alpha i}^2 \bm{x}_{\alpha i} \bm{x}_{\alpha i}^\top\] where $\D$ is of full rank and $n^{-1}\sum^n _{i=1} |v_{\alpha i}|^2 |\bm{x}_{ i}|^4 < \infty$. Furthermore, for any two correct models $\alpha_1,\alpha_2 \in \A_c$ such that $\alpha_1 \subset \alpha_2$, $$\trace (\DD^{-1} \bm{\Delta}_{\alpha_2 2} \DD^{-1} \bm{\Gamma}_{\alpha_2}) - \trace (\bm{\Delta}_{\alpha_1} ^{-1} \bm{\Delta}_{\alpha_1 2} \bm{\Delta}_{\alpha_1} ^{-1} \bm{\Gamma}_{\alpha_1}) > 0.$$
    \item For all models $\alpha \in \A$,
    \begin{enumerate}[label = (C.\arabic*)]
    \item The bootstrap FAMM estimator $\hat{\bm{\beta}}^* _{\alpha} = \bm{\beta}_\alpha + o_{p^*} (1)$.
    \item For all incorrect models, $\E_* \hat{\bm{\beta}}^* _\alpha - \hat{\bm{\beta}}_{\alpha} = o_p(1)$.
    \item For all correct models $\alpha \in \A_c$, for some $\bm{B}_\alpha,$ $m(\E_* \hat{\bm{\beta}}^* _\alpha - \hat{\bm{\beta}}_\alpha) = \bm{B}_\alpha + o_p(1)$.
    \item For all correct models $\alpha \in \A_c$ and scalar $\kappa$, $m\V_*(\hat{\bm{\beta}}^* _{\alpha}) = n\kappa\V (\hat{\bm{\beta}}_\alpha) + o_p(1)$.
\end{enumerate}
\end{enumerate}

With the addition of stratification indexation, the Conditions (C.1) -- (C.2) follow as an intermediate step in \citet[Proof of Theorem 1.]{salibian-barrera_robust_2008}. Condition (C.3) is established in \citet[Section 3]{muller_outlier_2005} for the WLS bootstrap estimator, with the special case of $w_{i} = 1$, corresponding to ordinary least squares (OLS). More generally, \citet{muller_outlier_2005}, recommend choosing $w_{i}$ to match the weights of a Mallows M-estimator. We extend this idea by taking the weights to be those of the full fit MM-estimator. That is, $w_{i} = v_{ i}$ for all $i \in \{1, \dots, n\}$. The validation of these conditions is provided in Appendix \ref{app:theory}.

Conditions (D), (E), and (F) do not explicitly restrict the bootstrap estimator; however we briefly discuss them for completeness. In Condition (D), the number of observations per bootstrap sample $m$ and the penalty function $\delta$ must satisfy $\delta(n) = o(n/m)$ and $m = o(n^{1/2})$. In finite samples, this does not require $m$ to be smaller than $\sqrt{n}$, but rather that $m$ grows asymptotically slower than $\sqrt{n}$. Conditions (E) and (F) restrict the model selection loss function $\rho$, and its derivative $\psi$.

\subsection{Robust COMBSS}

As outlined in both Section \ref{chap:preliminaries} and \cite[Algorithm 3, SubsetMapV2]{moka_combss_2024}, the optimal model for each size $k$ is the model that minimises the MSE. The resulting subset paths are non-robust, and a small number of outlying observations may alter the optimal subset at every model size. Although replacing the MSE with the robust criterion (\ref{eq:eMSC}) is conceptually straightforward, doing so would negate the computational advantage of COMBSS, as the repeated fitting of MM-estimators throughout the optimisation procedure becomes prohibitively expensive. To preserve the scalability of COMBSS, we instead use FAMM to estimate (\ref{eq:eMSC}), without forfeiting model selection consistency, as established in Section \ref{sec:theory}. Importantly, FAMM may replace the loss function in any BSS method for which an MM-estimator could be used, but at a fraction of the computational cost. In addition, scaling the data by the weights of the full MM-estimator such that $\tilde{\bm{X}}_{\alpha_f} = \bm{V}^{1/2} \bm{X}_{\alpha_f}$ and $\tilde{\bm{y}} = \bm{V}^{1/2}\bm{y}$ provides additional robustness. The resulting procedure is summarised in Algorithm \ref{alg:COMBSS-FAMM}.

\let\oldnl\nl
\SetNlSty{texttt}{(}{)}
\SetAlgoVlined
\begin{algorithm}
\KwIn{$\bm{X}_{\alpha_f}, \bm{y}, B, H, m,n, \rho, \delta(n), \bm{w},
\bm{t}_{\text{path}}$}
\tcp{Initialisation and Reweighting Step}
$\hat{\bm{\beta}} _{\alpha_f} \gets \text{Compute MM-estimation on } (\bm{X}_{\alpha_f}, \bm{y})$\;
$v_i \gets \text{Extract weights from IRLS for observations }i \in\{ 1,\dots,n\}$\;
$\bm{V} \gets \diag(v_1,\dots, v_n)$\;
$\bm{r}_{\alpha_f} \gets \bm{y} - \bm{X}_{\alpha_f} \hat{\bm{\beta}}_{\alpha_f}$\;
$\tilde{\bm{X}}_{\alpha_f} \gets \bm{V}^{1/2} \bm{X}_{\alpha_f}$\;
$\tilde{\bm{y}} \gets \bm{V}^{1/2} \bm{y}$\;
$q \gets \min\big(\mathrm{nrow}(\tilde{\bm{X}}_{\alpha_f}), \mathrm{ncol}(\tilde{\bm{X}}_{\alpha_f}) - 1\big)$\;
$\mathcal{M}_k \gets \emptyset \ \forall\ k\in \{1,\dots,q\}$\;
\vspace{1em}
\tcp{Build Candidate Models with COMBSS}
\For{$\bm{t} = (t_1, \dots, t_p)^\top \in \bm{t}_{\text{path}}$}{
    Let $t_{j_1}, t_{j_2}, \dots, t_{j_q}$ be the $q$ largest elements of $\bm{t}$ in decreasing order\;
    \For{$k = 1$ \KwTo $q$}{
        Let $\bm{s}_k \in \{0,1\}^p$ with non-zero elements at $j_1, \dots, j_k$\;
        $\mathcal{M}_k \gets \mathcal{M}_k \cup \{\bm{s}_k\}$\;
    }
}
\vspace{1em}
\tcp{Choose Optimal Models via FAMM (Algorithm~\ref{alg:FAMM})}
\For{$k = 1$ \KwTo $q$}{
    $\mathcal{A}_k \gets \mathrm{unique}(\mathcal{M}_k)$\;
    $\bm{s}^*_k \gets \text{Do FAMM}(\bm{X}_{\alpha_f} = \tilde{\bm{X}}_{\alpha_f},\ \bm{y} = \tilde{\bm{y}},B, \mathcal{A}_k, H, m, \rho, \delta(n), \bm{w},\bm{V}, \bm{r}_{\alpha_f})$\;
}
\KwRet{$\mathcal{M} = \{\bm{s}^*_1, \dots, \bm{s}^*_q\}$}\;
\caption{Robust COMBSS}\label{alg:COMBSS-FAMM}
\end{algorithm}

The model selection consistency results establish FAMM as a valid replacement for the bootstrap MM-estimator. Specifically, provided the true model $\alpha_0$ is contained in the candidate space, the FAMM-optimal model identifies the true model with probability tending to one. Consequently, for Robust COMBSS, if $\mathcal{A}_k$ contains $\alpha_0$ when $k = |\bm{\beta}_{\mathcal{S}}|$, then $\bm{s}_k^* = \alpha_0$ with probability tending to one. It follows that the best subset solution path $\mathcal{M}$ contains $\alpha_0$ with probability tending to one, so applying FAMM to $\mathcal{M}$ also selects $\alpha_0$ with probability tending to one.

\section{Simulation Study}\label{chap:simulations}
This section presents an empirical simulation study. We first assessed the suitability of FAMM as an approximation to the MM-estimator across a range of settings, benchmarking it against the FRB in Section \ref{sec:rms}. We then evaluated Robust COMBSS against COMBSS with FRB, and the non-robust COMBSS in Section \ref{sec:combss}. Finally, we applied the proposed methodology to a National Basketball Association (NBA) dataset in Section \ref{sec:NBA}.

The code for the FRB and MM-estimator was sourced from the FRBModelSelection and MASS R-packages respectively \citep{r_core_team_r_2026, salibian-barrera_robust_2008, venables_modern_2002}. All simulation code can be found in \url{https://github.com/MartinHuangR/Robust-COMBSS}. The COMBSS simulations were built on the code from \citet{moka_combss_2024}; however, we incorporated our own modifications and extensions. The majority of the simulations were computed on a machine with two Intel Xeon Gold 6148 CPUs and 768 GB RAM.

\subsection{Empirical Suitability of FAMM}\label{sec:rms}
We used the empirical probability of selecting the true model as a proxy for method performance. The empirical probability is the proportion of times the true model was selected across all unique datasets. Alongside probability, we also studied the computational speed of the methods. While we tried to keep a fair comparison between methods, it is worth noting that (a) there was no parallelisation used in the bootstrapping steps (only for running simultaneous simulations), and (b) the bulk of the FRB code was written in C, the MM-estimation in R, and FAMM in Rcpp \citep{balamuta_rcpp}.

Throughout this study, we used the robust model selection criterion, defined in (\ref{eq:eMSC}). The weights of the criterion were chosen to match the MM-estimator, as supported by Section \ref{sec:theory}. Following \citet{muller_outlier_2005}, the loss function $\rho$ was chosen as $\rho(z) = \min(z^2, b^2)$ with $b = 2$, and the penalty term $\delta(n)$ was $2\log(n)$. Furthermore, we set $B=100$ bootstrap samples. 

The rows $\bm{x}_i$ of the design matrix $\bm{X}$ were generated from $N(0,\bm{\Sigma})$, where $(\bm{\Sigma})_{ij} =  0.5^{|i-j|}$ was a covariance matrix with Toeplitz design. The coefficients of signal variables $\bm{\beta}_\mathcal{S}$ were $1$, with $|\bm{\beta}_{\mathcal{S}}|$ denoting the number of signal variables. Furthermore, all noise variables coefficients $\bm{\beta}_{\mathcal{N}}$ were $0$. The response variable was computed with the model $y_i = \bm{x}^\top _{i} \bm{\beta} + \epsilon_i$, with $\bm{\beta} = [\bm{\beta}_{\mathcal{S}} ^ \top, \bm{\beta}_{\mathcal{N}}^\top ] ^\top$ and $\epsilon_i \sim N(0, \sigma^2)$. All simulations were also held at a signal-to-noise ratio (SNR) of 2, that is, $\lVert \bm{X} \bm{\beta}\rVert^2 _2 / n\sigma^2 = 2$.

To investigate the performance of the methods with outliers in the response and rows, we utilised a contamination model following \citet{christidis_robust_2026} and \citet{maronna_2011}. We contaminated $\nu$ rows, corresponding to a proportion $\tau = \nu/n$ of the sample, such that rowwise outliers $\bm{x}_i ^\dagger$ replaced the uncontaminated predictors $\bm{x}_i$ with the function
$$\bm{x}^\dagger_i = \bm{\Theta}_i + \frac{k_{\text{lev}}}{\sqrt{\bm{a}^\top \bm{\Sigma}^{-1} \bm{a}}}\bm{a},\ 1\leq i \leq \nu,$$ where $\bm{\Theta}_i \sim N(\bm{0}_{p},\ 0.01\times \bm{I}_p)$ and $\bm{a} = \bm{a}^\dagger- (1/p)\bm{a}^{\dagger\top} \bm{1}_{p}$. Here, $\bm{I}_p$ is a $p$-dimensional identity matrix, $\bm{0}_{p} = (0,\dots,0)^\top \in \mathbb{R}^p$, and $\bm{1}_{p} = (1,\dots,1)^\top \in \mathbb{R}^p$. The elements $a^\dagger_j$ in $\bm{a}^\dagger$ followed $U(-1,1)$ for $1\leq j\leq p$. The response variable $y_i$ was contaminated by misleading the regression coefficient $\bm{\beta}$ by a factor of $(1 + k_{\text{slo}})$. The contaminated response variable was computed as $y^\dagger_{i} = \bm{x}^{\dagger\top} _i \bm{\beta}^\dagger,\ 1 \leq i \leq \nu,$ where the $j$th entry in $\bm{\beta}^\dagger$ was
\begin{align}
\beta^\dagger_j =
\begin{cases}
    \beta_j(1 + k_{\text{slo}}),\ &\beta_j \neq 0,\\
    k_{\text{slo}}\lVert\bm{\beta}\rVert_\infty,\ &\text{otherwise,}
\end{cases}  \ 1\leq j\leq p.
\end{align}

The two parameters $k_{\text{lev}}$ and $k_{\text{slo}}$ controlled the leverage and slope of the contamination respectively. Following \citet{christidis_robust_2026}, we set $k_{\text{lev}} = 2$ and $k_{\text{slo}} = 100$. Under this contamination model, we assessed the proportions $\tau\in \{0,0.05,0.1,0.15,0.2\}$.

We present two simulation settings, (i) $n = 500,\ p = 50,\ |\bm{\beta}_\mathcal{S}| = 3$ and (ii) $n =1000,\ p = 100,\ |\bm{\beta}_\mathcal{S}| = 10$. For each setting, we generated $1000$ unique datasets, used $H=8$ strata, and set bootstrap sample sizes of $m = 250$ and $500$ respectively. An exhaustive search of the model space was not feasible due to computational limitations. Therefore, we chose the candidate set to include the true model (that is, the indices of the variables with a non-zero coefficient), two underfitted models, and two overfitted models.

We defined the least underfitted model as the model that included all signal variables except the one with the smallest absolute coefficient. The second least underfitted model excluded the two signal variables with the smallest absolute coefficients. We also defined an overfitted model as the model that included all signal variables together with the noise variable that had the highest correlation with the response. The second overfitted model included all signal variables and the two noise variables with the highest correlations with the response.

\begin{figure}
    \centering
    \includegraphics[scale = 0.5]{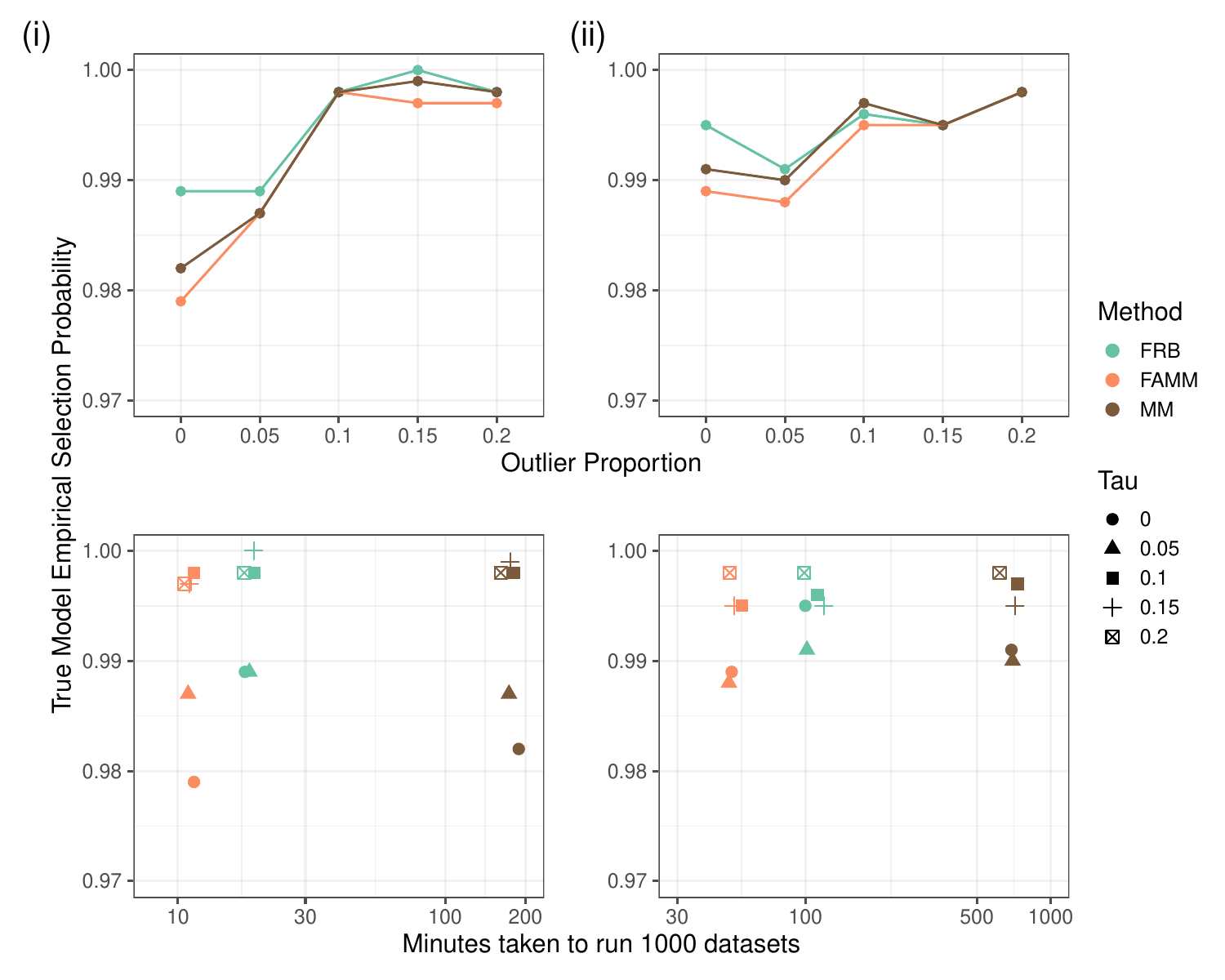}
    \caption{True model $\alpha_0$ empirical selection probability for artificial data with settings (i) $n = 500,\ p = 50,\ |\bm{\beta}_\mathcal{S}| = 3$ and (ii) $n =1000,\ p = 100,\ |\bm{\beta}_\mathcal{S}| = 10$.}
    \label{fig:S14}
\end{figure}

The results in Figure \ref{fig:S14} empirically support the claim that the FAMM estimator is an appropriate approximation to the MM-estimator, with nearly identical true model selection probabilities. FAMM exhibits the fastest computation speed out of the three methods, being roughly $2$ times faster than FRB and $16$ times faster than MM in setting (i), and $2$ times faster than FRB and $14$ times faster than MM in setting (ii).

In further analyses not shown for brevity, we observed that increasing the sample size of the dataset while holding the number of predictors constant allowed the true model selection probability to increase across all contamination levels. Furthermore, increasing the number of predictors while keeping the sample size fixed did not drastically affect performance, but rather led to a substantial increase in computational time. Across all simulation designs, a consistent trend was identified: FRB, FAMM, and MM all had similar model performance, with FAMM being substantially faster.

For completeness, we also include the benchmark dataset used in both \citet{muller_outlier_2005} and \citet{salibian-barrera_robust_2008}, the results of which can be found in Appendix \ref{app:bench}. 

\subsection{Robust COMBSS}\label{sec:combss}

In this section, we controlled the output of COMBSS to be a single optimal model per model size $k \in \{1,\dots, p-1\}$. To analyse performance, we introduce the definition of a ``correct'' model. When $k < |\bm{\beta}_{\mathcal{S}}|$, a correct model is any subset of the true model. Conversely, if $k > |\bm{\beta}_{\mathcal{S}}|$, a correct model is any superset of the true model. When $k = |\bm{\beta}_{\mathcal{S}}|$, a correct model must be the true model.

We considered two simulated settings: (i) $n = 200,\ p = 20,\ |\bm{\beta}_{\mathcal{S}}|=3$ and (ii) $n = 500,\ p = 50,\ |\bm{\beta}_{\mathcal{S}}| = 5$ with a fixed SNR of 2, and contamination levels $k_{\text{slo}} = 100$ and $k_{\text{lev}} = 2$. For each setting we generated $1000$ unique datasets and set $m = 100$ and $250$ respectively, with both containing $H=8$ strata. As in the previous example, the coefficients of signal variables were $1$, whereas noise variables had coefficients of $0$. The criterion also had a weighting structure matching the MM-estimator weights. We benchmarked Robust COMBSS against COMBSS with FRB and the non-robust COMBSS.

\begin{figure}
    \centering
    \includegraphics[scale = 0.4]{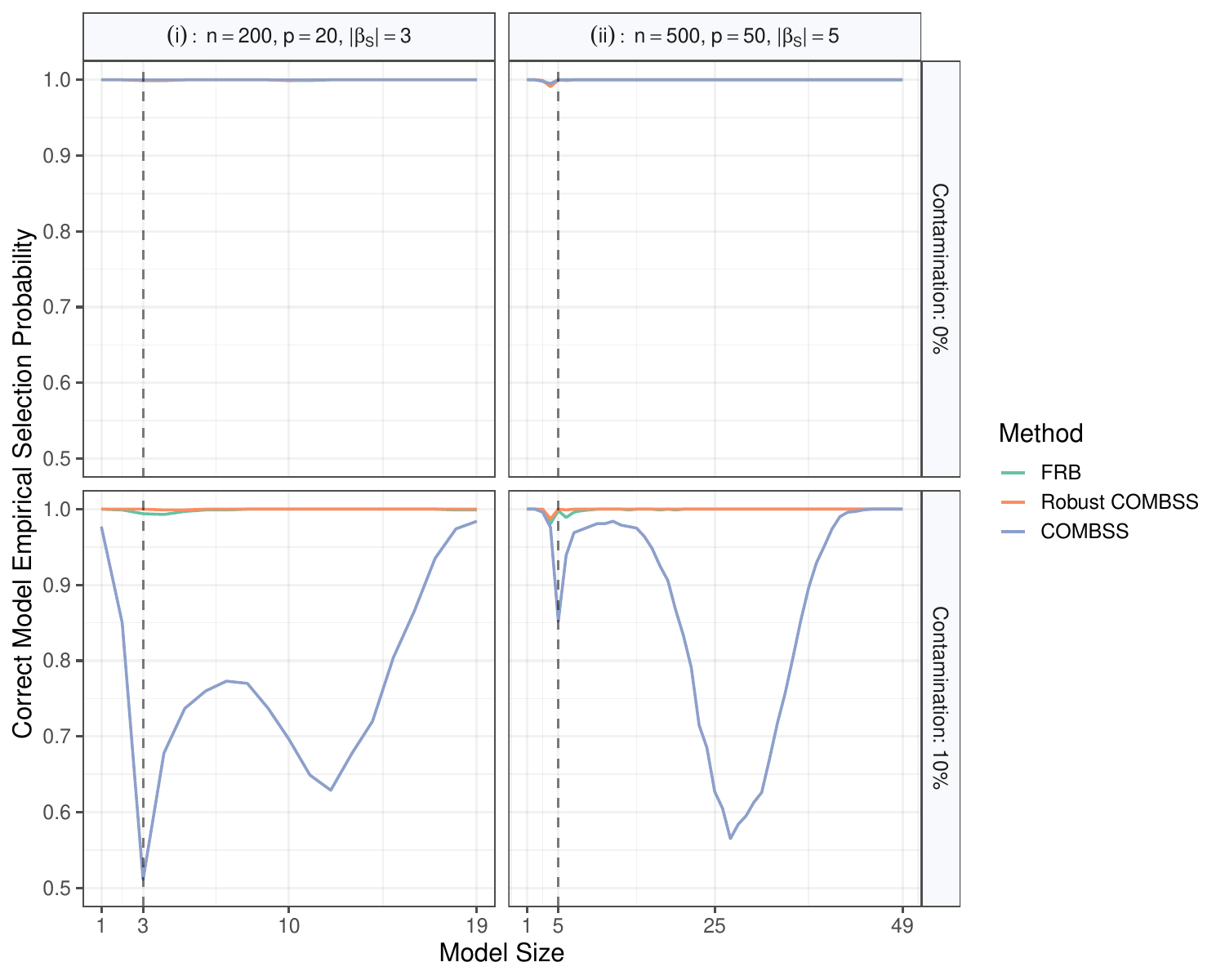} 
    \caption{Correct model empirical selection probability per model size for artificial data with settings (i) $n =200,\ p = 20,\ |\bm{\beta}_\mathcal{S}| = 3$ and (ii) $n = 500,\ p = 50,\ |\bm{\beta}_\mathcal{S}| = 5$ for 0\% and 10\% contamination. The dashed vertical line indicates $|\bm{\beta}_{\mathcal{S}}|$, when a correct model is required to be a true model.}
    \label{fig:C}
\end{figure}

\begin{figure}
    \centering
    \includegraphics[scale = 0.4]{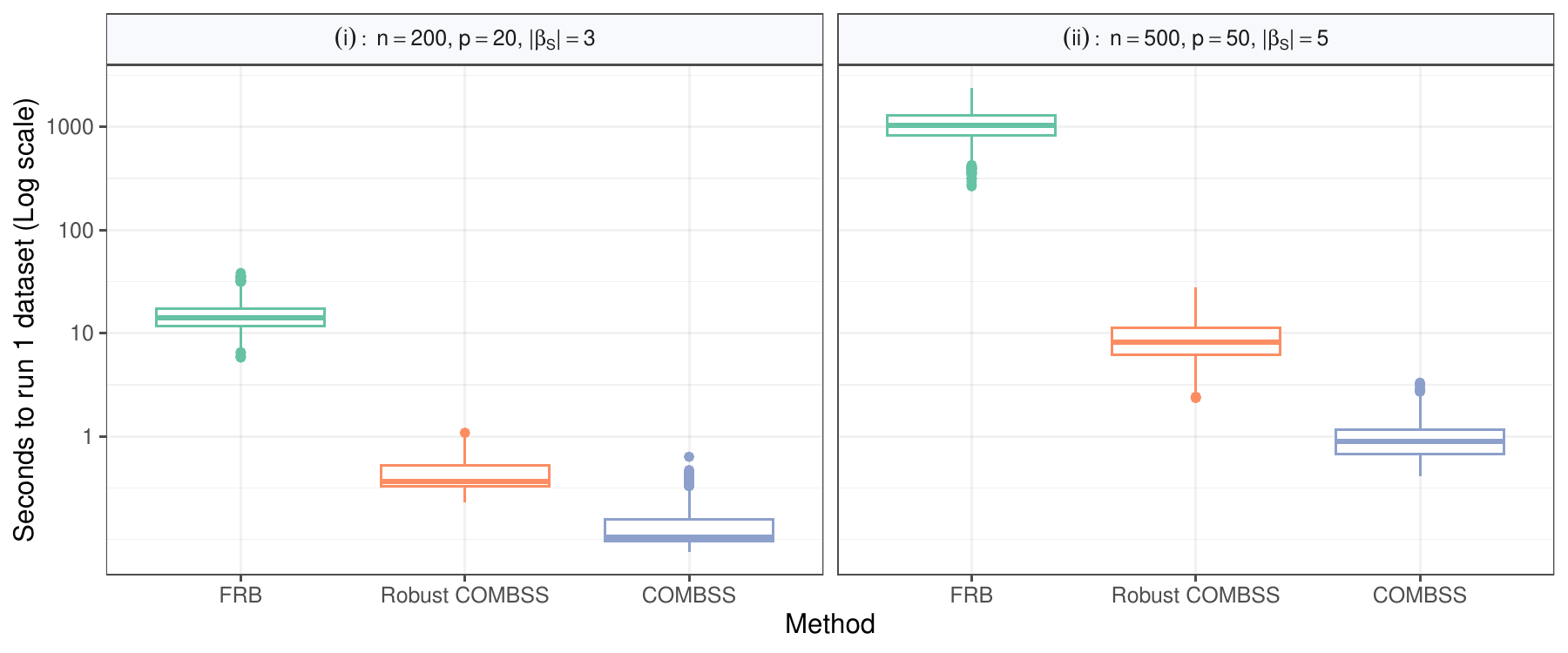} 
    \caption{Seconds to run 1 dataset (Log scale) with COMBSS for settings (i) $n =200,\ p = 20,\ |\bm{\beta}_\mathcal{S}| = 3$ and (ii) $n = 500,\ p = 50,\ |\bm{\beta}_\mathcal{S}| = 5$.}
    \label{fig:Ct}
\end{figure}

When the data are uncontaminated in settings (i) and (ii), the model selection probabilities are almost perfect in all methods, as shown in Figure \ref{fig:C}. In the 10\% contamination case, there is a clear drop-off in performance for COMBSS, while the two robust methods remain largely unaffected.  

It is expected that there is a dip in selection probabilities when the model size equals the true number of signal variables (vertical dashed line), as a correct model can only be the true model. Furthermore, as the model size increases, the probability of selecting the correct model also increases, which explains why COMBSS gradually improves its correct model selection probabilities as the model size approaches $p$ in the 10\% contamination case across both settings (i) and (ii).

While Robust COMBSS outperforms the non-robust counterpart, it is much slower, as expected, shown in Figure \ref{fig:Ct}. Specifically, Robust COMBSS requires $3$ and $9$ times the computation time of COMBSS in settings (i) and (ii), respectively. However, the substantial increase in performance justifies the additional computation time. Conversely, Robust COMBSS is considerably faster than COMBSS with FRB, being $35$ times faster for (i) and $120$ times faster in (ii).

We next benchmarked our methodology with the simulation settings and parameter configurations used in \citet{moka_combss_2024} with code given in \url{https://github.com/benoit-liquet/COMBSS-R-VIGNETTE/blob/main/Low_dimensional_example.md}. This simulation study examined the effect of Robust COMBSS in the context of determining an optimal model per regularisation parameter $\lambda$, as outlined in \citet[Algorithm 2., SubsetMapV1]{moka_combss_2024}. Since SubsetMapV1 does not use a specific criterion, the difference between Robust COMBSS and the non-robust COMBSS is the rescaling of data through the MM-estimator, as specified in Section \ref{chap:rCOMBSS}. 

Apart from the dataset dimensions, the simulation settings were identical to those used previously. Since FAMM is not applicable in the high-dimensional setting ($p > n$), we considered only low-dimensional scenarios. Outlier contamination was introduced using the same model and parameter settings described in Section \ref{sec:rms}. We considered two settings: (i) $n = 100,\ p = 20,\ |\bm{\beta}_{\mathcal{S}}| = 10$  and (ii) $n = 1000,\ p = 100,\ |\bm{\beta}_{\mathcal{S}}| = 10$.
\begin{figure}
    \centering
    \includegraphics[scale = 0.3]{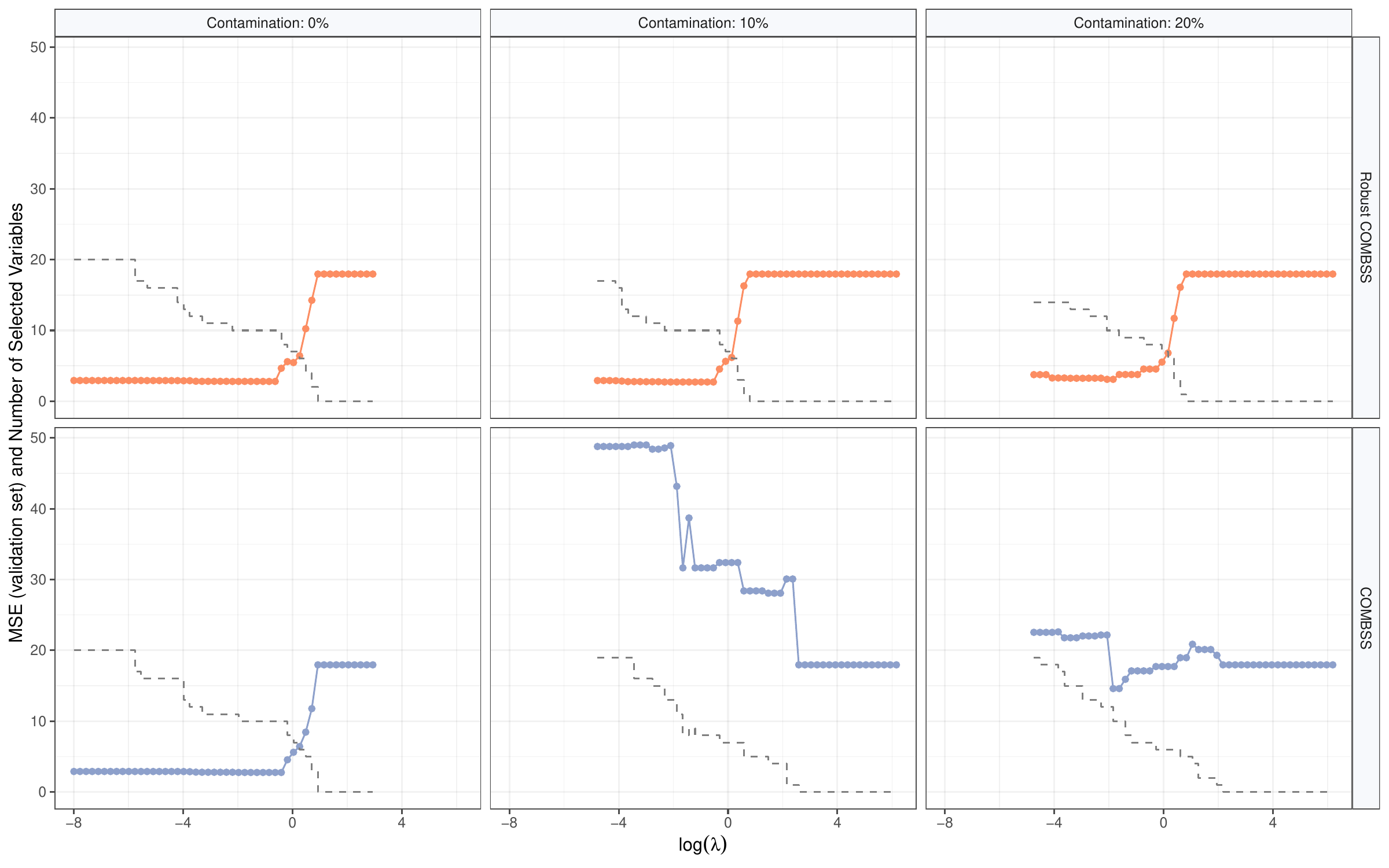} 
    \caption{Regularisation path for setting (i) $n = 100,\ p= 20,\ |\bm{\beta}_{\mathcal{S}}| = 10$ with 0\%, 10\%, and 20\% contamination. Coloured line denotes validation set MSE. Dashed line denotes the number of selected variables.}
    \label{fig:bcombss1}
\end{figure}

\begin{figure}
    \centering
    \includegraphics[scale = 0.3]{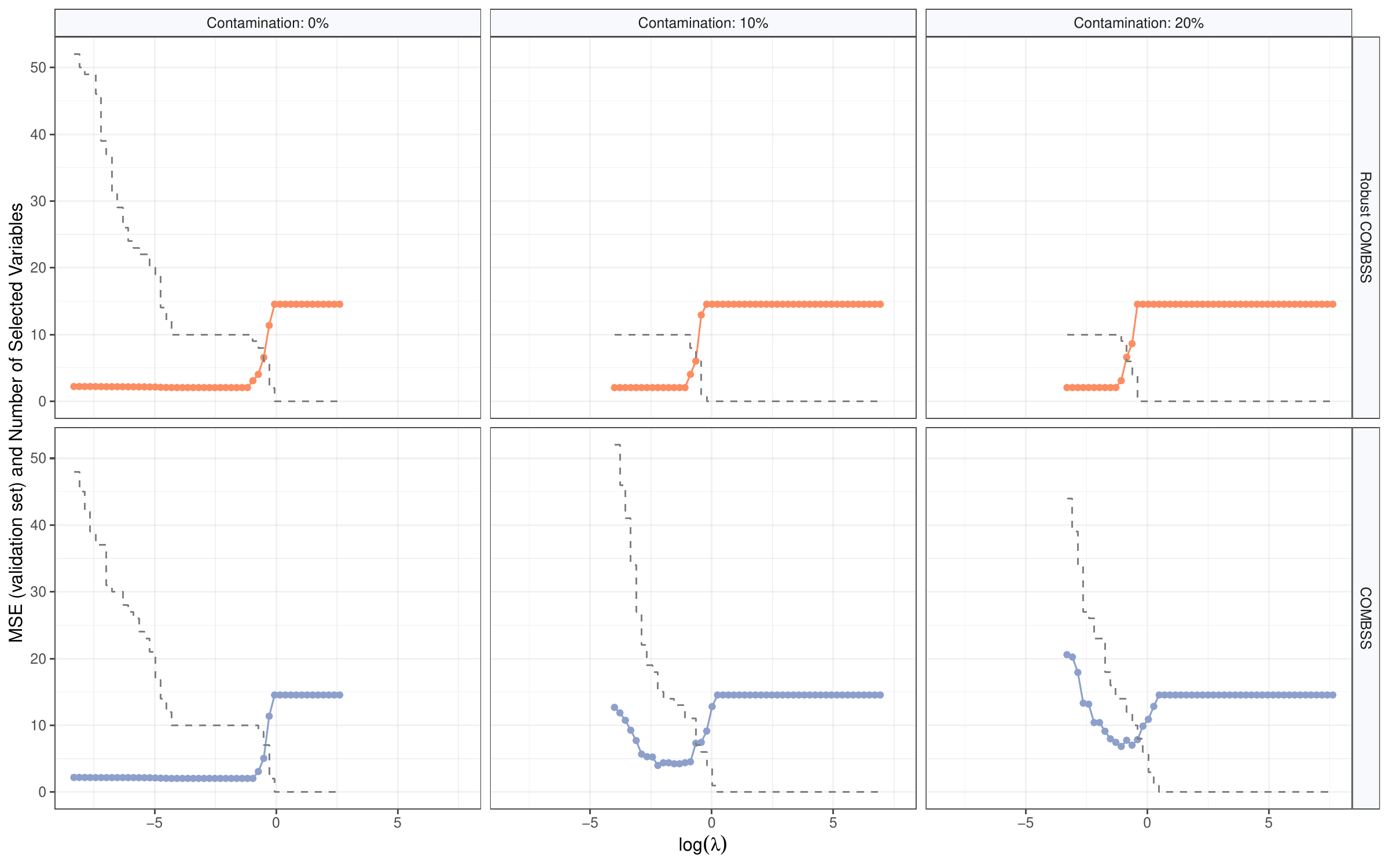} 
    \caption{Regularisation path for setting (ii) $n = 1000,\ p= 100,\ |\bm{\beta}_{\mathcal{S}}| = 10$ with 0\%, 10\%, and 20\% contamination. Coloured line denotes validation set MSE. Dashed line denotes the number of selected variables.}
    \label{fig:bcombss2}
\end{figure}

In both settings, presented in Figures \ref{fig:bcombss1} and \ref{fig:bcombss2}, the introduction of outliers substantially increases the validation set MSE for the non-robust COMBSS. Conversely, the Robust COMBSS is largely unaffected by contamination. Furthermore, in both contamination scenarios and across both settings, Robust COMBSS achieves a lower validation set MSE while selecting substantially smaller models. Notably, Robust COMBSS does not exhibit an increase in MSE under the non-contaminated setting, suggesting that it is an appropriate default choice when the proportion of outliers is unknown.

\subsection{National Basketball Association Data}\label{sec:NBA}
This section applied an exhaustive search and COMBSS to a dataset from the 2024/2025 National Basketball Association (NBA) regular season. 

Position-based outliers in the NBA typically arise when players exhibit playstyles outside traditional norms, such as three-point scoring centers, or rebounding point guards. While point guards have traditionally led the league in assist percentage, recent years have seen a notable increase in assist rates among forwards and centers as well. These patterns are driven by a small subset of taller players with outlying assist percentages relative to the typically low averages for players of a similar height. Recent literature has deemed traditional positional labels ambiguous, suggesting the use of height-based classification, or the redefinition of roles, such as low assist percentage role-playing bigs, or playmaking versatile forwards \citep{muniz_weighted_2022, south_basketball_2025, wang_will_2023}. Accordingly, we utilise the height-based classification approach and focus on players whose heights fall in the top quartile ($\geq$ 205.74 cm, 6 ft 9 in). This threshold lies just below one standard deviation of the mean height of the tallest and heaviest performance profile cluster identified by \citet{zhang_clustering_2018}.

We constructed our data using two datasets of size $n = 567,\ p = 21$ and $n = 569,\ p = 23$, where each observation was an individual player. The datasets were merged by player names, and variables were removed based on irrelevance and/or multicollinearity issues. We then subset the data to contain players with a height greater than or equal to $205.74$ cm, or $6$ ft $9$ in. As per-game averages can be misrepresented by a small game sample size, we only analysed players who participated in at least $41$ games -- half the total number of regular season games. Our final dataset had a size of $n = 95$ and $p = 12$, where the assist percentage was the response variable. Some explanatory variables included height, weight, true shooting and usage rate. The design matrix was scaled such that it had a mean of $0$ and standard deviation of $1$. We then constructed our candidate models $\mathcal{A}$. As $p$ was small, we were able to conduct an exhaustive search. Assuming all models include an intercept term, we had $2^{12} = 4096$ total models. We also set $m = 48$ and $H=8$. Data availability, metadata, and exploratory data analysis are provided in Appendix \ref{app:nba}.

\begin{figure}
    \centering
    \includegraphics[scale = 0.25]{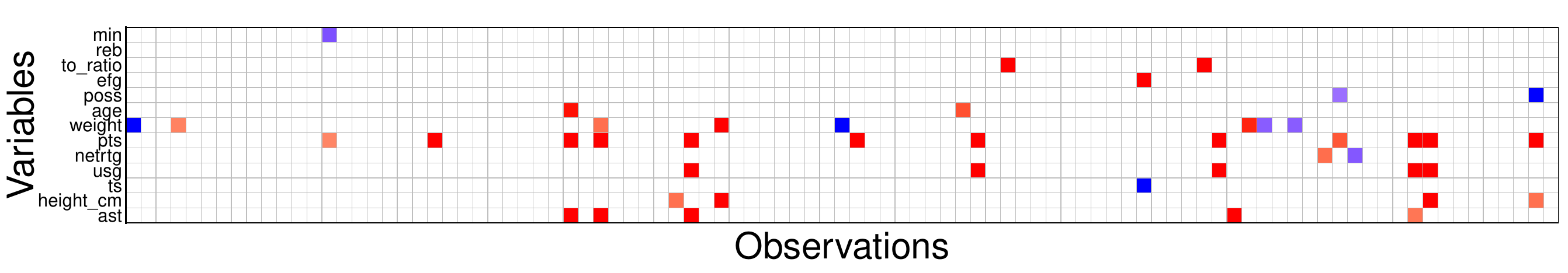} 
    \caption{Deviating cells map of NBA data. Red cells indicate positive outliers where observed values exceed predictions, blue cells indicate severe negative outliers below predictions, and purple cells indicate milder outliers with moderate deviations.}
    \label{fig:DDC}
\end{figure}

To visualise the proportion of cell-wise outliers in the dataset, we used the Detect Deviating Cells method (DDC), shown in Figure \ref{fig:DDC} \citep{rousseeuw_detecting_2018}. The cell map highlights estimated cell-wise outliers by comparing their standardised residuals to their expected value. DDC indicates the data has an average of approximately 4\% outliers per predictor variable. Around 28\% of rows have at least one outlying observation. Points (pts) has the highest percentage of outliers at 13\%, while reb (rebound percentage) has none.

We applied FAMM, FRB and the MM robust bootstrap to the dataset, with parameter settings identical to previous simulations. For a non-robust comparison, we also included the Akaike Information Criterion (AIC) \citep{akaike_information_1998}. The optimal model from $\mathcal{A}$ was then selected using the robust model selection criterion in (\ref{eq:eMSC}). Once identified, a 10 repeated 5-fold cross-validation was conducted to estimate the model's average mean absolute error (MAE). All computations were executed in Rcpp, except for FRB, which was in C.

\begin{table}
\centering
\resizebox{\textwidth}{!}{%
\begin{tabular}{cccc}
\hline
\textbf{Method} 
                & \textbf{MAE (SD)}
                & \textbf{Computation Time (s)}
                & \textbf{Model Chosen} \\
\hline
FRB  & 4.37 (0.64)& 91.78    & usg, reb \\
FAMM & 4.42 (0.67)  & 3.56    & usg \\
MM   & 4.42 (0.67) & 5527.25 & usg \\
AIC  & 4.37 (0.75) & 5.38    &  reb, min, to\_ratio, efg, age, pts, ts, height\\
\hline
\end{tabular}}
\caption{10 repeated 5-fold cross-validation mean absolute error (MAE), standard deviation (SD), and model selection computation time on NBA data for an exhaustive search with 4096 candidate models.}
\label{tab:NBAres}
\end{table}

Since the true model is unknown, we cannot compare model selection accuracy. However, the model selected by FAMM in Table \ref{tab:NBAres} is identical to that obtained by MM. Not only does FAMM select the same model as MM, but it also reduces the computation time of the MM from over an hour and a half to under five seconds, corresponding to a $1550$ times speedup.

The usage (usg) variable selected by FAMM and MM in Table \ref{tab:NBAres} has a positive association with the response variable assist percentage. Usage rate refers to the percentage of team plays used by a player when they are on the floor. This has a direct impact on how many assists a player has. If a player is not involved in a play, they cannot gain an assist. The three highest usage rate players in the 2024/2025 season all exhibited assist percentages in the $95$th percentile.

Although FAMM and MM yield marginally higher MAE than the FRB and AIC selected models, the differences are negligible, and within one standard deviation of each other. The substantially larger model chosen by AIC offers little predictive gain while carrying greater risks of overfitting. The model chosen by FRB is similar to FAMM and MM, differing by only a rebound (reb) variable. The inclusion of rebounds to predict assist percentage is difficult to justify, as a player's rebounding has little conceptual bearing on their playmaking. This suggests that the variable may be helpful for predictive purposes, but is not a meaningful explanatory factor. While there is a slight advantage in MAE, the FRB is slower than FAMM by $26$ times.

Overall, the above results indicate that FAMM provides an effective surrogate for the MM-estimator, producing equivalent model selection results while delivering a substantial computational speedup. This reduction in computational cost makes FAMM particularly attractive in settings involving large numbers of predictors or candidate models.

\begin{figure}
    \centering
    \includegraphics[scale = 0.4]{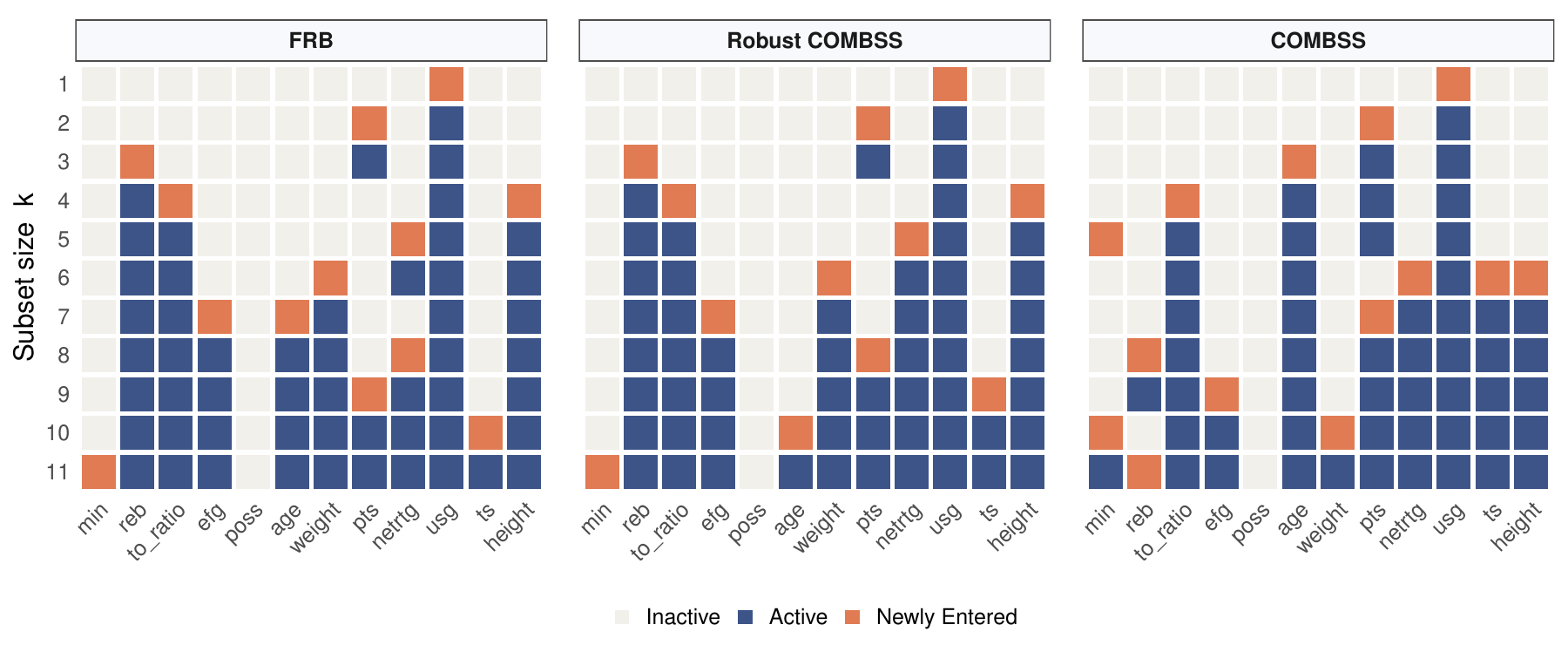} 
    \caption{Best subset solution path with COMBSS on NBA data. Variables re-entering the solution path indicate an unstable path caused by outliers.}
    \label{fig:bpath}
\end{figure}

We then fit COMBSS on the NBA dataset to produce a best subset solution path. We used identical parameter settings in the COMBSS framework, as specified previously. The three methods compared were Robust COMBSS, COMBSS with FRB and the non-robust COMBSS. 

All methods select the same model for $k=1$, with usg as the sole predictor, as shown in the solution path in Figure \ref{fig:bpath}. Furthermore, Robust COMBSS identifies the same model as in the exhaustive search in Table \ref{tab:NBAres}. Although FRB selects a model with two variables in the exhaustive search, this model does not appear in the solution path for $k = 2$. In all methods, poss is not selected, aligning with the final models in Table \ref{tab:NBAres}. Under the non-robust COMBSS, several predictors (min, reb, pts) enter and drop out of the active set across subset sizes, indicating an unstable solution path disrupted by outliers. This is highlighted at $k = 6$, when three variables enter the path at once. In the Robust COMBSS, this re-entering of variables only happens in pts.

\section{Conclusion}\label{chap:conclusion}
Best subset selection produces candidate models for each model size $k$, but its reliance on a squared error loss leaves the resulting selection vulnerable to even a small number of outliers. While exact BSS is intractable beyond a small number of predictor variables, COMBSS overcomes this by relaxing the discrete search into a continuous problem, making solution paths feasible for larger numbers of predictor variables. A natural way to make BSS robust replaces the non-robust loss with a robust model selection criterion estimated via an MM-estimator. This extension, however, is difficult to scale. Even though COMBSS makes BSS computationally cheaper, refitting the MM-estimator for every candidate model across a large candidate space is still computationally prohibitive. To address this issue, we introduced FAMM, a fast approximate MM-estimator to enable Robust COMBSS for contaminated data.

FAMM provides a simple and computationally efficient approximation to the MM-estimator in the robust stratified bootstrap model selection procedure. Since the observations are down-weighted through a single full-data MM-estimator, the resulting model selection criterion remains robust while being substantially faster to compute. As FAMM uses fixed weights over bootstrap samples, it is no longer an MM-estimator, meaning model selection consistency properties do not automatically transfer. We demonstrated that FAMM is indeed model selection consistent when the model selection weights are set equal to the full data MM-estimator weights. Empirically, we showed that FAMM matches the model selection performance of the MM-estimator while achieving a substantial reduction in computation time. 

FAMM is primarily intended for model selection rather than post-selection inference. The fixed weights compromise precise estimates of coefficient uncertainty, and are less suitable for tasks such as confidence interval construction. From a model selection perspective, however, this is not restrictive, as the resulting loss of variance does not affect the asymptotic probability of selecting the true model. In applications where post-selection inference is required, FAMM may be used to efficiently identify the model, after which a full MM-estimator can be refit on the selected model for final coefficient estimation and confidence interval construction. While our analysis is focused primarily on COMBSS, FAMM can be readily integrated into other BSS methods, as it simply substitutes the loss function. 

Overall, FAMM provides a practical balance between computational efficiency and robustness in bootstrap model selection. By substantially reducing computation time without sacrificing model selection performance, it makes robust BSS feasible in settings where refitting an MM-estimator would otherwise be prohibitive. Embedded within COMBSS, FAMM extends this advantage to large predictor spaces, delivering a simple and practical procedure for robust BSS at scale.

\section*{Acknowledgements}

Samuel Muller and Garth Tarr were supported by the Australian Research Council (DP260100348).

\clearpage
\bibliographystyle{apalike} 
\bibliography{ref.bib}

\clearpage
\begin{appendices}
\section{Theoretical Results}\label{app:theory}
This appendix gives additional details into the conditions as outlined in Section \ref{sec:theory}.

\subsection{Condition (C.1)}
Following the steps in \citet[Pg. 5132]{salibian-barrera_robust_2008} and rewriting in stratification form,
\begin{align*}
    \hat{\bm{\beta}}^* _{\alpha} - \hat{\bm{\beta}}_{\alpha} &= \left[\sum^H _{h=1}\sum^{m_h}_{i=1} v_{hi}^* \bm{x}_{\alpha hi}^* \bm{x}_{\alpha  hi}^{*\top} \right]^{-1} \sum^H _{h=1}\sum^{m_h}_{i=1} v_{hi}^* \bm{x}^*_{\alpha hi} y_{hi}^* - \hat{\bm{\beta}}_{\alpha}\\
    &= \left[\frac{1}{m}\sum^H _{h=1} \sum^{m_h} _{i=1} v^* _{hi} \bm{x}^* _{\alpha hi} \bm{x}^{*\top} _{\alpha hi}\right]^{-1} \left[\frac{1}{m}\sum^H _{h=1} \sum^{m_h} _{i=1} \psi \left(\frac{y^* _{hi} - \bm{x}^{*\top}_{\alpha h i} \hat{\bm{\beta}}_{\alpha}}{\hat{\sigma}}\right)\bm{x}^{*\top}_{\alpha h i} \right]\\
    &\eqqcolon \bm{A}^{* -1}_m \bm{b}^* _m\\
    &= o_{p^*}(1),
\end{align*}
and therefore $\hat{\bm{\beta}}^* _\alpha = \hat{\bm{\beta}}_\alpha + o_{p^*} (1)$. From \citet{yohai_high_1987}, $\hat{\bm{\beta}}_{\alpha} - \bm{\beta}_{\alpha} = o_p (1)$. It follows that
\begin{align*}
    \hat{\bm{\beta}}^* _{\alpha} - \bm{\beta}_\alpha 
    &= (\hat{\bm{\beta}}^* _{\alpha} - \hat{\bm{\beta}}_{\alpha}) + (\hat{\bm{\beta}}_{\alpha} - \bm{\beta}_{\alpha})\\
    &= o_{p^*} (1)+ o_p(1)\\
    &= o_{p^*}(1)\\
    \Rightarrow\hat{\bm{\beta}}^* _{\alpha} &= \bm{\beta}_{\alpha} + o_{p^*} (1).
\end{align*}
\subsection{Condition (C.2)}
Similar to Condition (C.1), we follow the same steps as \citet{salibian-barrera_robust_2008}. Assigning the selection criterion weights as the same as the MM-estimator,
\begin{align*}
\lim _{n \rightarrow \infty} \bm{A}_m ^* =  \frac{1}{m} \sum^H _{h=1}\sum^{m_h} _{i=1} w^* _{hi} \bm{x}^* _{\alpha hi} \bm{x}^{*\top} _{\alpha hi} \rightarrow^p \bm{\Gamma}^* _\alpha  > 0.
\end{align*}
Then,
\begin{align*}
    \bm{A}^{*-1} _m \bm{b}^* _m &= \bm{A}^{*-1} _m\bm{b}^* _m + (\bm{\Gamma}^* _\alpha)^{-1} \bm{b}^* _m - (\bm{\Gamma}^* _\alpha)^{-1} \bm{b}^* _m \\
    &= \left[\bm{A}^{*-1} _m -(\bm{\Gamma}^* _\alpha)^{-1} \right]\bm{b}^* _m + (\bm{\Gamma}^* _\alpha)^{-1}\bm{b}^*_m.
\end{align*}
Since $\E_* \bm{b}^* _m  = \bm{0}$, $\bm{A}^*_m$ and $\bm{\Gamma}^* _\alpha$ are non-singular, and $\bm{A}^*_m = \bm{\Gamma}^* _\alpha + (\bm{A}^*_m - \bm{\Gamma}^* _\alpha)$, we have
\begin{align*}
     \bm{A}^{*-1} _m = (\bm{\Gamma}^* _\alpha)^{-1} - (\bm{\Gamma}^* _\alpha)^{-1}(\bm{A}^* _m - \bm{\Gamma}^* _\alpha)(\bm{I} + (\bm{\Gamma}^* _\alpha)^{-1}(\bm{A}^*_m - \bm{\Gamma}^* _\alpha))^{-1}(\bm{\Gamma}^* _\alpha)^{-1}.
\end{align*}
If $\bm{A}^* _m - \bm{\Gamma}^* _\alpha = o_p(1)$ and $\bm{A}^{*-1} _m - (\bm{\Gamma}^* _\alpha)^{-1} = o_p (1)$,
\begin{align*}
    \bm{A}^{*-1} _m \bm{b}^* _m &= \left[\bm{A}^{*-1} _m -(\bm{\Gamma}^* _\alpha)^{-1} \right]\bm{b}^* _m + (\bm{\Gamma}^* _\alpha)^{-1}\bm{b}^*_m\\
    &= o_p(1) + (\bm{\Gamma}^* _\alpha)^{-1}\bm{b}^*_m
\end{align*}
Taking the expectation, $\E_* [\bm{A}^{*-1} _m \bm{b}^* _m] =  o_p(1)$. Since $\hat{\bm{\beta}}^* _{\alpha} - \hat{\bm{\beta}}_{\alpha} = \bm{A}^{* -1} _m \bm{b}^* _m$ as defined in Condition (C.1), $\E_* (\hat{\bm{\beta}}^* _{\alpha} - \hat{\bm{\beta}}_{\alpha}) = \E_* [\bm{A}^{*-1} _m \bm{b}^* _m] = o_p (1)$, and therefore $\E_*\hat{\bm{\beta}}^* _{\alpha} - \hat{\bm{\beta}}_{\alpha} = o_p(1)$ for all models. 

\subsection{Condition (C.3)}
Recall for the bootstrap estimator, the weights are treated as fixed and do not depend on the model index $\alpha$. For brevity we suppress the $\alpha$ subscript.

From Condition (A), let the $p_\alpha \times p_{\alpha}$ matrix $\bm{\Gamma}_{\alpha} = \lim_{n \rightarrow \infty} n^{-1} \sum^n _{i=1} w_{i}\bm{x}_{\alpha i} \bm{x}_{\alpha i}^\top$ be full rank, and $n^{-1} \sum^n _{i=1} |w_{i}|^2 |\bm{x}_i |^4 < \infty$. Also, denote $\bm{1}_{hi}$ as an indicator function, such that $\bm{1}_{hi} = \bm{1}(r_{\alpha n, [nq_{h-1}]+1} \leq r_{\alpha i} \leq r_{\alpha n, [n q_h] + 1})$. From \citet{muller_outlier_2005}, we know that 
\begin{align*}
    \sum^H _{h=1} \sum^{m_h} _{i=1} w_{hi}^* \bm{x}^* _{\alpha hi} \bm{x}_{\alpha hi}^{*\top} &= \sum^H _{h=1} \frac{m_h }{n_h} \sum^{n}_{i=1} w_{hi} \bm{x}_{\alpha i}\bm{x}_{\alpha i}^\top \bm{1}_{hi} \{1 + o_p (1)\} \\
    &= \frac{m}{n} \sum^n _{i=1} w_i \bm{x}_{\alpha i} \bm{x}_{\alpha i}^\top \{1 + o_p (1)\}\\
    &= m\bm{\Gamma}_{\alpha}\{1 + o_p (1)\},\ \text{where } \bm{\Gamma}_{\alpha} = \lim_{n \rightarrow \infty} n^{-1} \sum^n _{i=1} w_{i}\bm{x}_{\alpha i} \bm{x}_{\alpha i}^\top.
\end{align*}
Here, $\sum^H _{h=1} \sum^{m_h} _{i=1} w_{hi} \bm{x}^* _{\alpha hi} \bm{x}_{\alpha hi}^\top$ is asymptotically non-singular and $\bm{\Gamma}_\alpha$ is also non-singular and invertible. The bootstrap estimator is
\begin{align*}
    \hat{\bm{\beta}}^* _{\alpha} = \left(\sum^H _{h=1} \sum^{m_h}_{i=1} w^* _{hi} \bm{x}^* _{\alpha hi} \bm{x}^{*\top}_{\alpha hi}\right)^{-1} \sum^H _{h=1} \sum^{m_h}_{i=1} w^* _{hi}  \bm{x}^* _{\alpha hi} y_{hi}^*.     
\end{align*}
With $y^* _{hi} = \bm{x}^{*\top} _{\alpha hi} \hat{\bm{\beta}}_{\alpha} + y^* _{hi} - \bm{x}^{*\top} _{\alpha hi} \hat{\bm{\beta}}_{\alpha}$,
\begin{align*}
    \hat{\bm{\beta}}^* _{\alpha} &= \left(\sum^H _{h=1} \sum^{m_h}_{i=1} w^* _{hi} \bm{x}^* _{\alpha hi} \bm{x}^{*\top}_{\alpha hi}\right)^{-1} \sum^H _{h=1} \sum^{m_h}_{i=1} w^* _{hi}  \bm{x}^* _{\alpha hi} (\bm{x}^{*\top} _{\alpha hi} \hat{\bm{\beta}}_{\alpha} + y^* _{hi} - \bm{x}^{*\top} _{\alpha hi} \hat{\bm{\beta}}_{\alpha}) \\
    &= \hat{\bm{\beta}}_{\alpha}+ \left(\sum^H _{h=1} \sum^{m_h}_{i=1} w^* _{hi} \bm{x}^* _{\alpha hi} \bm{x}^{*\top}_{\alpha hi}\right)^{-1}\sum^H _{h=1} \sum^{m_h}_{i=1} w^* _{hi}  \bm{x}^* _{\alpha hi}(y^* _{hi} - \bm{x}^{*\top} _{\alpha hi} \hat{\bm{\beta}}_{\alpha})\\
    &\eqqcolon  \hat{\bm{\beta}}_\alpha + \bm{A}^{*-1} \bm{\omega}^*
\end{align*}
where $\bm{A}^* \coloneqq \left(\sum^H _{h=1} \sum^{m_h}_{i=1} w^* _{hi} \bm{x}^* _{\alpha hi} \bm{x}^{*\top}_{\alpha hi}\right)$ and $\bm{\omega}^* \coloneqq \sum^H _{h=1} \sum^{m_h}_{i=1} w^* _{hi}  \bm{x}^* _{\alpha hi}(y^* _{hi} - \bm{x}^{*\top} _{\alpha hi} \hat{\bm{\beta}}_{\alpha})$. 
Writing $\bm{A}^{*-1} = \bm{A}^{*-1} + m^{-1} \bm{\Gamma}^{-1} _{\alpha} - m^{-1} \bm{\Gamma}^{-1} _{\alpha}$ and substituting,
\begin{align}\label{eq:bootstrapeq}
    \hat{\bm{\beta}}^*_{\alpha} &= \hat{\bm{\beta}}_{\alpha} + (\bm{A}^{*-1} + m^{-1} \bm{\Gamma}^{-1} _{\alpha} - m^{-1} \bm{\Gamma}^{-1} _{\alpha})\bm{\omega}^*\nonumber\\
    &= \hat{\bm{\beta}}_{\alpha}+ m^{-1} \bm{\Gamma}_{\alpha}^{-1} \bm{\omega}^* - m^{-1}\bm{\Gamma}_{\alpha} ^{-1}\left(\sum^H _{h=1} \sum^{m_h}_{i=1} w^* _{hi} \bm{x}^* _{\alpha hi} \bm{x}^{*\top}_{\alpha hi} - m\bm{\Gamma}_{\alpha}\right) \nonumber\\
    &\qquad \times m^{-1}\bm{\Gamma}^{-1}_{\alpha}\bm{\omega}^*\{1 + o_p (1)\},
\end{align}
using the fact that $\bm{A}^{*-1} - m^{-1}\bm{\Gamma}_\alpha^{-1} = -m^{-1}\bm{\Gamma}_\alpha^{-1}(\bm{A}^* - m\bm{\Gamma}_\alpha)\bm{A}^{*-1}$.

We will now consider the expectation of the bootstrap estimator. Using the fact that $\frac{1}{n}\sum^n _{i=1} \psi \left(\frac{y_i - \bm{x}^\top _{\alpha i}\hat{\bm{\beta}}_{\alpha}}{\hat{\sigma}}\right)\bm{x}_{\alpha i} = \bm{0}$ and $\frac{m}{n}\sum^n _{i=1} w_i\bm{x}_{\alpha i}  (y_i - \bm{x}^\top _{\alpha i}\hat{\bm{\beta}}_{\alpha}) = \bm{0}$ we have that $\E_* \bm{\omega}^* = 0$. Taking the expectation of the bootstrap estimator yields
\begin{align*}
    \E_* \hat{\bm{\beta}}^* _\alpha &= \hat{\bm{\beta}}_{\alpha} - m^{-2} \bm{\Gamma}^{-1} _\alpha \E_* \left[ \sum^H _{h=1} \sum^{m_h}_{i=1} w^* _{hi} \bm{x}^* _{\alpha hi} \bm{x}^{*\top}_{\alpha hi}\bm{\Gamma}_\alpha ^{-1} \sum^H _{h=1}\sum^{m_h}_{i=1}w^* _{hi} \bm{x}^* _{\alpha hi} (y^* _{hi} - \bm{x}^{*\top} _{\alpha hi} \hat{\bm{\beta}}_\alpha)\right]\{1 + o_p (1)\}\\
    &= \hat{\bm{\beta}}_\alpha - m^{-2} \bm{\Gamma}^{-1}_\alpha \sum^H _{h=1} \sum^{m_h} _{i=1}\E_* \left[ w^* _{hi} \bm{x}^* _{\alpha hi} \bm{x}^{*\top}_{\alpha hi}\bm{\Gamma}_\alpha ^{-1} w^* _{hi} \bm{x}^* _{\alpha hi} (y^* _{hi} - \bm{x}^{*\top} _{\alpha hi} \hat{\bm{\beta}}_\alpha)\right] \{1 + o_p (1)\} \\
    & \qquad \qquad + m^{-2}\bm{\Gamma}^{-1}_\alpha \sum^H _{h=1} \sum^{m_h} _{i=1} \sum^H _{\substack{h' = 1 \\(h,i) \neq (h',i')}} \sum^{m_h} _{\substack{i' = 1 \\(h,i) \neq (h',i')}}  \E_* \left[w^* _{hi} \bm{x}^* _{\alpha hi} \bm{x}^{*\top} _{\alpha hi}  \right]\\
    & \qquad \qquad \qquad \qquad \times \E_* \left[w^* _{h'i'} \bm{x}^* _{\alpha h'i'} (y^* _{h'i'} - \bm{x}^{*\top}_{\alpha h'i'}\hat{\bm{\beta}}_\alpha) \right]\{1 + o_p(1)\}
\end{align*}
The off-diagonal term can be expanded as
\begin{align}
    &m^{-2}\bm{\Gamma}^{-1}_\alpha\left(\sum_{h=1}^H \sum^{m_h}_{i=1}\E_* \left[w^* _{hi} \bm{x}^* _{\alpha hi} \bm{x}^{*\top} _{\alpha hi} \right] \sum^H _{h'=1}\sum^{m_{h'}}_{i'=1} \E_* \left[w^* _{h'i'} \bm{x}^* _{\alpha h'i'} (y^* _{h'i'} - \bm{x}^{*\top}_{\alpha h'i'}\hat{\bm{\beta}}_\alpha)\right]\right.\nonumber \\
     &\qquad \qquad- \left.\sum^H _{h=1} \sum^{m_h} _{i=1} \E_* \left[w^* _{hi} \bm{x}^* _{\alpha hi} \bm{x}^{*\top} _{\alpha hi}  \right] \E_* \left[w^* _{hi} \bm{x}^* _{\alpha hi} (y^* _{hi} - \bm{x}^{*\top}_{\alpha hi}\hat{\bm{\beta}}_\alpha) \right]\right)\{1+o_p(1)\}\label{eq:off1}\\
    &=  o_p(m^{-1})\nonumber
\end{align}
It then follows that 
\begin{align*}
   \E_* \hat{\bm{\beta}}^* _\alpha &= \hat{\bm{\beta}}_\alpha - m^{-2} \bm{\Gamma}^{-1}_\alpha \sum^H _{h=1} \sum^{m_h} _{i=1}\E_* \left[ w^{*2} _{hi} \bm{x}^* _{\alpha hi} \bm{x}^{*\top}_{\alpha hi}\bm{\Gamma}_\alpha ^{-1}  \bm{x}^* _{\alpha hi} (y^* _{hi} - \bm{x}^{*\top} _{\alpha hi} \hat{\bm{\beta}}_\alpha)\{1 + o_p (1)\}\right] +o_p(m^{-1}) \\
    &= \hat{\bm{\beta}}_{\alpha } - m^{-1}n^{-1} \bm{\Gamma}^{-1} _{\alpha} \sum^n _{i=1} w_{i}^2 \bm{x}_{\alpha i}\bm{x}_{\alpha i}^\top \bm{\Gamma}^{-1}_{\alpha}  \bm{x}_{\alpha i} \epsilon_{\alpha i}\{1 + o_p (1)\} + o_p(m^{-1}).
\end{align*}
By the law of large numbers $n^{-1} \sum^n _{i=1} w_{i}^2\bm{x}_{\alpha i} \bm{x}^\top _{\alpha i} \bm{\Gamma}^{-1} _\alpha\bm{x}_{\alpha i} \epsilon_{\alpha i} \rightarrow ^p \E\left[w_{i}^2\bm{x}_{\alpha i} \bm{x}^\top _{\alpha i} \bm{\Gamma}^{-1} _\alpha \bm{x}_{\alpha i} \epsilon_{\alpha i}  \right]$. Furthermore, for all models, $\bm{\epsilon}_{\alpha} = \bm{\epsilon} + \bm{x}^\top _{\alpha^c i} \bm{\beta}_{\alpha^c}$ \citep{muller_outlier_2005}. Equivalently, $\epsilon_{\alpha i} = \epsilon_i + \bm{x}^\top _{\alpha^c i} \bm{\beta}_{\alpha^c}\ \forall i \in \{1,\dots,n\}$. Then
\begin{align*}
    \E\left[w_{i}^2\bm{x}_{\alpha i} \bm{x}^\top _{\alpha i} \bm{\Gamma}^{-1} _\alpha \bm{x}_{\alpha i} \epsilon_{\alpha i}  \right] 
    &= \E\left[w_{i}^2\bm{x}_{\alpha i} \bm{x}^\top _{\alpha i} \bm{\Gamma}^{-1} _\alpha \bm{x}_{\alpha i} \epsilon_i\right] + \E\left[w_{i}^2\bm{x}_{\alpha i} \bm{x}^\top _{\alpha i} \bm{\Gamma}^{-1} _\alpha\bm{x}_{\alpha i} \bm{x}^\top _{\alpha^c i}\bm{\beta}_{\alpha ^ c}\right]\\
    &= \E\left[w_{i}^2\bm{x}_{\alpha i} \bm{x}^\top _{\alpha i} \bm{\Gamma}^{-1} _\alpha \bm{x}_{\alpha i} \bm{x}^\top _{\alpha^c i}\bm{\beta}_{\alpha ^ c}\right]
\end{align*}
as $\E\left[w_{i}^2\bm{x}_{\alpha i} \bm{x}^\top _{\alpha i} \bm{\Gamma}^{-1} _\alpha \bm{x}_{\alpha i} \epsilon_i\right] = w_{i}^2\bm{x}_{\alpha i} \bm{x}^\top _{\alpha i} \bm{\Gamma}^{-1} _\alpha\bm{x}_{\alpha i} \E\left[\epsilon_i\right] = 0$, since $\bm{x}_{\alpha i}$, $w_{i}$, and $\bm{\Gamma}^{-1} _{\alpha}$ are non-random and $\E[\epsilon_i]$ is $0$. Let the matrix
\begin{align*}
    \bm{B}_\alpha &= \lim_{n\rightarrow \infty} -n^{-1} \bm{\Gamma}^{-1} _{\alpha} \sum^n _{i=1} w_{i}^2 \bm{x}_{\alpha i}\bm{x}_{\alpha i}^\top \bm{\Gamma}^{-1}_{\alpha} \bm{x}_{\alpha i} \epsilon_{\alpha i} \\
    &= -\bm{\Gamma}^{-1} _\alpha \E\left[w_{i}^2\bm{x}_{\alpha i} \bm{x}^\top _{\alpha i} \bm{\Gamma}^{-1} _\alpha\bm{x}_{\alpha i} \bm{x}^\top _{\alpha^c i}\bm{\beta}_{\alpha ^ c}\right]\\
    &= \bm{0},
\end{align*}
since $\bm{\beta}_{\alpha^c} = \bm{0}$ for correct models. Then,
\begin{align*}
    \E_* \hat{\bm{\beta}}^* _\alpha &= \hat{\bm{\beta}}_{\alpha } - m^{-1}n^{-1} \bm{\Gamma}^{-1} _{\alpha} \sum^n _{i=1} w_{i}^2 \bm{x}_{\alpha i}\bm{x}_{\alpha i}^\top \bm{\Gamma}^{-1}_{\alpha} \bm{x}_{\alpha i} \epsilon_{\alpha i}\{1 + o_p (1)\} + o_p(m^{-1})\\
    m(\E_* \hat{\bm{\beta}}^* _{\alpha} - \hat{\bm{\beta}}_{\alpha})  &= -\bm{\Gamma}^{-1}_{\alpha} n^{-1}\sum^n _{i=1} w_{i}^2 \bm{x}_{\alpha i}\bm{x}_{\alpha i}^\top \bm{\Gamma}^{-1}_{\alpha}  \bm{x}_{\alpha i} \epsilon_{\alpha i}\{1 + o_p (1)\} + o_p(1)\\
    &= \bm{B}_{\alpha} + o_p(1)
\end{align*}

\subsection{Condition (C.4)}
We establish this condition for $\kappa = 1$, which implies that the bootstrap variance and the sampling variance agree up to the factor $m/n$. 

Consider the previously computed bootstrap estimator in (\ref{eq:bootstrapeq}). We will first show that the first term $m^{-1} \bm{\Gamma}_{\alpha}^{-1} \bm{\omega}^*$ dominates the equation. Conditional on the bootstrap sample, the sums are independent across strata and i.i.d within each stratum with $\E_* \bm{\omega}^*= 0$. Since $\bm{\omega}^*$ is a sum of $m$ conditionally independent centered terms with bounded conditional second moments, $\bm{\omega}^*= O_{p^*}(\sqrt{m})$. It follows that $$m^{-1} \bm{\Gamma}^{-1} _\alpha \bm{\omega}^*= O_{p^*} (m^{-1/2}).$$

Now we will focus on the second term 
\begin{align}\label{eq:seconds}
  &m^{-1}\bm{\Gamma}_{\alpha} ^{-1}\left(\sum^H _{h=1} \sum^{m_h}_{i=1} w^* _{hi} \bm{x}^* _{\alpha hi} \bm{x}^{*\top}_{\alpha hi} - m\bm{\Gamma}_{\alpha}\right)m^{-1}\bm{\Gamma}^{-1}_{\alpha}\bm{\omega}^*\{1 + o_p (1)\}.
\end{align}
Firstly, $\left(\sum^H _{h=1} \sum^{m_h}_{i=1} w^* _{hi} \bm{x}^* _{\alpha hi} \bm{x}^{*\top}_{\alpha hi} - m\bm{\Gamma}_{\alpha}\right) = o_p(m)$, using the fact that $\sum^H _{h=1} \sum^{m_h} _{i=1} w_{hi}^* \bm{x}^* _{\alpha hi} \bm{x}_{\alpha hi}^{*\top} = m\bm{\Gamma}_{\alpha}\{1 + o_p (1)\}.$ Then (\ref{eq:seconds}) is a remainder term of order $o_{p^*}(m^{-1/2})$. Therefore we are left with 
\begin{align*}
    \hat{\bm{\beta}}^*_\alpha &= \hat{\bm{\beta}}_\alpha + m^{-1}\bm{\Gamma}_\alpha ^{-1} \bm{\omega}^*+ R
\end{align*}
where $R \coloneqq  o_{p^*}(m^{-1/2})$. Taking the variance gives
\begin{align*}
    \V_*(\hat{\bm{\beta}}^*_\alpha) &= \V_*(m^{-1} \bm{\Gamma}^{-1}_\alpha \bm{\omega}^*) + \V_*(R) + 2\cov_*(m^{-1}\bm{\Gamma}^{-1}_\alpha \bm{\omega}^*, R).
\end{align*}
The covariance can be evaluated as
\begin{align*}
    2\cov_*(m^{-1}\bm{\Gamma}^{-1}_\alpha \bm{\omega}^*, R) &= o_p (m^{-1}).
\end{align*}
Therefore,
\begin{align}\label{eq:V} 
    \V_* (\hat{\bm{\beta}}^* _{\alpha}) &= m^{-2} \bm{\Gamma}_{\alpha}^{-1} \V_* \left(\bm{\omega}^*\right) \bm{\Gamma}^{-1} _\alpha + o_{p}(m^{-1}).
\end{align}
Let $r^* _{\alpha hi} = y^* _{hi} - \bm{x}^{*\top}_{\alpha hi} \hat{\bm{\beta}}_{\alpha}$. Following the steps in (\ref{eq:off1}),
\begin{align*}
    \V_* (\bm{\omega}^*)&=\E_* \left[\bm{\omega}^*\bm{\omega}^{*\top} \right]\\
     &= \E_*\left[\sum^H _{h=1} \sum^{m_h} _{i=1} w^{2*}_{hi} \bm{x}^* _{\alpha hi} \bm{x}^{*\top}_{\alpha hi} r^{2*} _{\alpha hi} \right] \\
    & \qquad \qquad + \sum^H _{h=1} \sum^{m_{h}} _{i=1} \sum^H _{\substack{h' = 1 \\(h,i) \neq (h',i')}}\sum^{m_{h'}} _{\substack{i' = 1 \\(h,i) \neq (h',i')}}\E_*\left[w^* _{hi}  \bm{x}^* _{\alpha hi}r^* _{\alpha hi}\right]  \E_*\left[ w^* _{h'i'}  \bm{x}^{*\top} _{\alpha h'i'}r^* _{\alpha h'i'}\right]\\
    &= \E_*\left[\sum^H _{h=1} \sum^{m_h} _{i=1} w^{2*}_{hi} \bm{x}^* _{\alpha hi} \bm{x}^{*\top}_{\alpha hi} r^{2*} _{\alpha hi} \right] + o_p(m).
\end{align*}
From Condition (C.3) we know that 
\begin{align*}
     \sum^H _{h=1} \sum^{m_h} _{i=1} w^{2*}_{hi} \bm{x}^* _{\alpha hi} \bm{x}^{*\top}_{\alpha hi} r^{2*} _{\alpha hi}  = \frac{m}{n}\sum^n _{i=1} w^2 _i\bm{x}_{\alpha i}\bm{x}_{\alpha i}^\top r^2 _{\alpha i}\{1 + o_p(1)\},
\end{align*}
which leads to 
\begin{align*}
    \V_* (\bm{\omega}^*) &= \E_*\left[\frac{m}{n}\sum^n _{i=1} w^2 _i\bm{x}_{\alpha i}\bm{x}_{\alpha i}^\top r^2 _{\alpha i}\{1 + o_p(1)\} \right] + o_p(m) \\ 
    &= mn^{-1}\sum^n _{i=1} w^2 _i\bm{x}_{\alpha i}\bm{x}_{\alpha i}^\top r^2 _{\alpha i}\{1 + o_p(1)\}.
\end{align*}
Subbing this back to (\ref{eq:V}), we get
\begin{align*}
    \V_* (\hat{\bm{\beta}}^* _{\alpha}) &= m^{-1} n^{-1} \bm{\Gamma}^{-1}_\alpha \sum^n _{i=1} w^2 _i\bm{x}_{\alpha i}\bm{x}_{\alpha i}^\top r^2 _{\alpha i}\{1 + o_p(1)\} \bm{\Gamma}^{-1} _\alpha + o_{p}(1).
\end{align*}
From \citet{muller_outlier_2005}, we know that Condition (B) is true for an MM-estimator. Therefore, we can say that $n\V(\hat{\bm{\beta}}_\alpha) = \phi \bm{\Delta}^{-1} _\alpha \DD \bm{\Delta}^{-1}_\alpha + o_p(1)$, where $\bm{\Delta}_\alpha = \lim_{n\rightarrow \infty} n^{-1} \sum^n _{i=1} v_{\alpha i} \bm{x}_{\alpha i}\bm{x}_{\alpha i}^\top$ and $\DD = \lim_{n\rightarrow \infty} n^{-1} \sum^n _{i=1} v_{\alpha i}^2 \bm{x}_{\alpha i}\bm{x}_{\alpha i}^\top$. Since we have assumed that the weights of a MM-estimator are to be set as the weights in the model selection criterion, we can see that $\bm{\Delta}_\alpha$ is indeed equal to $\bm{\Gamma}_\alpha$, and their inverse is also equal. Furthermore, we can rewrite $\DD$ as $\DD = \lim_{n\rightarrow \infty} n^{-1} \sum^n _{i=1} w_{i}^2 \bm{x}_{\alpha i}\bm{x}_{\alpha i}^\top$. Therefore, showing $m \V_* (\hat{\bm{\beta}}^* _{\alpha}) = n\kappa \V (\hat{\bm{\beta}}_\alpha) + o_p(1)$ is true (Condition (C.4)) follows by taking $\kappa = 1$ and showing 
\begin{align*}
m \V_* (\hat{\bm{\beta}}^* _{\alpha}) = \phi \bm{\Delta}^{-1} _\alpha \DD \bm{\Delta}^{-1}_\alpha + o_p(1).
\end{align*}
Substituting $\bm{\Gamma}_\alpha$ for $\bm{\Delta}_\alpha$, 
$$m\V_* (\hat{\bm{\beta}}^* _{\alpha}) = \bm{\Delta}^{-1}_\alpha n^{-1} \sum^n _{i=1} w^2_i \bm{x}_{\alpha i} \bm{x}^\top _{\alpha i} r^2_{\alpha i} \{1 + o_p(1)\}\bm{\Delta}^{-1}_\alpha+ o_{p}(1).$$
Since $r_{\alpha i} =  y_i - \bm{x}^\top _{\alpha i}\bm{\beta}_\alpha + \bm{x}_{\alpha i}^\top (\bm{\hat{\beta}}_{\alpha} - \bm{\beta}_{\alpha}) = \epsilon_{\alpha i} + O_p(n^{-1/2})$, it follows that $r^2_{\alpha i} = \epsilon^2 _{\alpha i} + O_p(n^{-1/2}).$ Then,
\begin{align*}
    n^{-1} \sum^n _{i=1} w^2_i \bm{x}_{\alpha i} \bm{x}^\top _{\alpha i} r^2_{\alpha i} &= n^{-1} \sum^n _{i=1} w^2_i \bm{x}_{\alpha i} \bm{x}^\top _{\alpha i} (\epsilon^2 _{\alpha i} + O_p(n^{-1/2}))\\
    &= \sigma^2 \DD + o_p(1).
\end{align*}
Therefore,
\begin{align*}
    m\V_* (\hat{\bm{\beta}}^* _\alpha)  &= \bm{\Delta}_{\alpha}^{-1} (\sigma^2 \DD + o_p(1))\bm{\Delta}_\alpha ^{-1} \{1 + o_p(1)\} + o_{p}(1)\\
    &= \phi\bm{\Delta}^{-1}_\alpha \DD\bm{\Delta}^{-1}_\alpha + o_{p}(1),\ \ \phi \coloneqq \sigma^2.
\end{align*}
The right-hand side is precisely $n\V(\hat{\bm{\beta}}_\alpha) + o_p(1)$ from Condition (B), so
$$m\V_*(\hat{\bm{\beta}}^*_\alpha) = n\kappa \V(\hat{\bm{\beta}}_\alpha) + o_p(1)$$
when $\kappa = 1$, verifying Condition (C.4).

\newpage
\section{Benchmarked Simulation}\label{app:bench}
We benchmarked our method with the simulation setting used by \citet{muller_outlier_2005}, and later \citet{salibian-barrera_robust_2008}. A total of 1000 unique datasets were generated from $y_i = 2 + 2x_{i1} + 0x_{i2} + \epsilon_i,\ i = 1,\dots,64$, where the components of the design matrix $\bm{X}$ was generated from a uniform distribution $U(0,1)$. There were six different types of error distributions $F_\epsilon$ considered.

\begin{enumerate}[label=(\roman*)]
\item N: The standard normal distribution $N(0,1)$
\item 1/8: A normal mixture distribution $7/8 N(0,1) + 1/8N(30-2-2x_1, 1)$
\item 1/4: A normal mixture distribution $3/4N(0,1) + 1/4N(30 - 2 - 2x_1, 1)$
    \item 3/8: A normal mixture distribution $5/8 N(0,1) + 3/8N(30-2-2x_1, 1)$
    \item C: The Cauchy distribution $C(0,1)$
    \item S: The slash distribution $N(0,1)/U(0,1)$
\end{enumerate}

Each dataset had $B = 100$ bootstrap samples of size $m = 24$. For both FAMM and MM-estimation, the stratified bootstrap was employed with $H=8$ strata. With $\alpha_f =\{1,2\}$, the candidate model space $\A$ included the true model $\alpha_0 = \{1\}$, and $\alpha_1 = \emptyset, \alpha_2 = \{2\}, \alpha_3 = \{1, 2\}$. All candidate models included the intercept term $\hat{\beta}_0$. 
 
\begin{figure}[h]
    \centering
    \includegraphics[scale = 0.4]{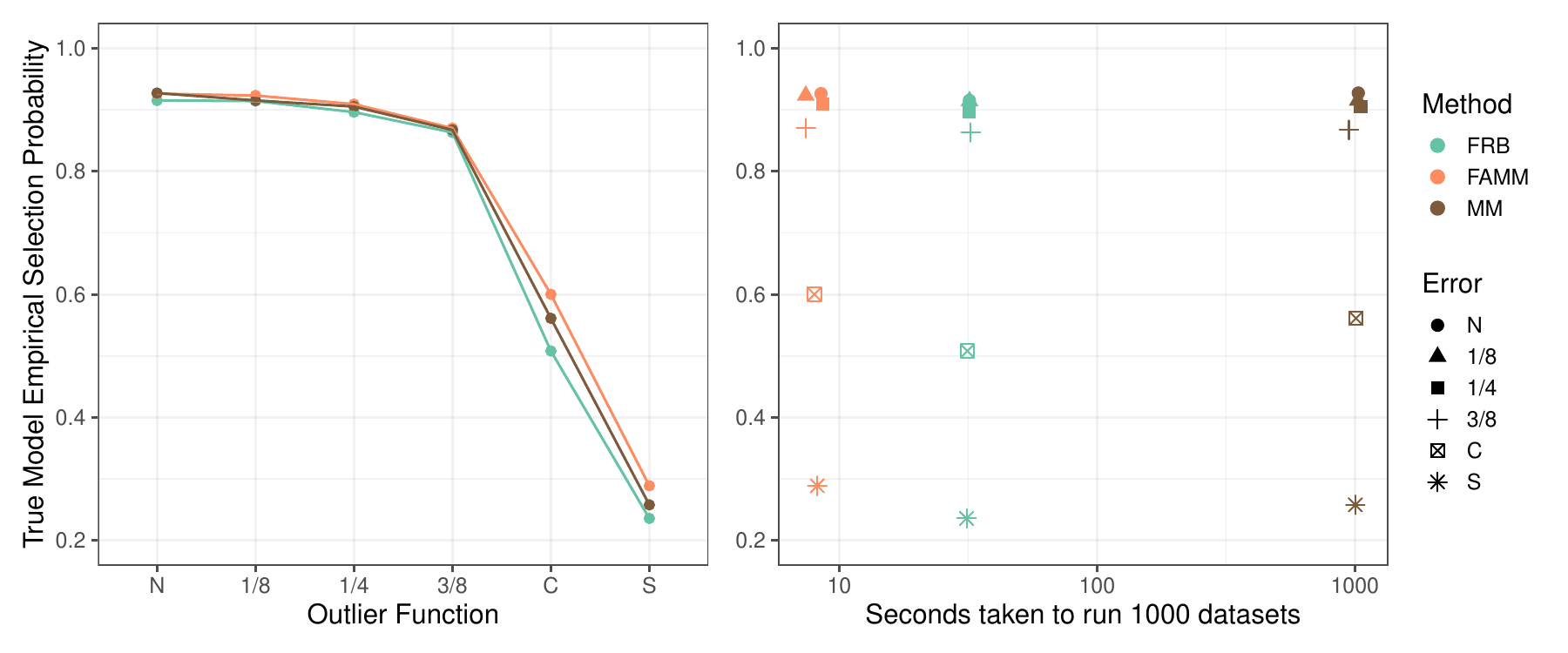} 
    \caption{True model $\alpha_0$ empirical selection probability for artificial data with settings (i) - (vi).}
    \label{fig:SM}
\end{figure}

Figure \ref{fig:SM} highlights that FRB, FAMM, and MM methods perform comparably under the standard normal and normal mixture error distributions. However, none exhibits encouraging results for the Cauchy and slash distributions. In these cases, the FRB method slightly underperforms relative to FAMM and MM. Notably, FAMM consistently matches the performance of MM, highlighting its suitability as an approximation. Furthermore, FAMM achieves an average speedup of $4$ times over FRB and $126$ times over MM.

\newpage
\section{NBA Exploratory Analysis}\label{app:nba}
\begin{figure}
    \centering
    \includegraphics[scale = 0.4]{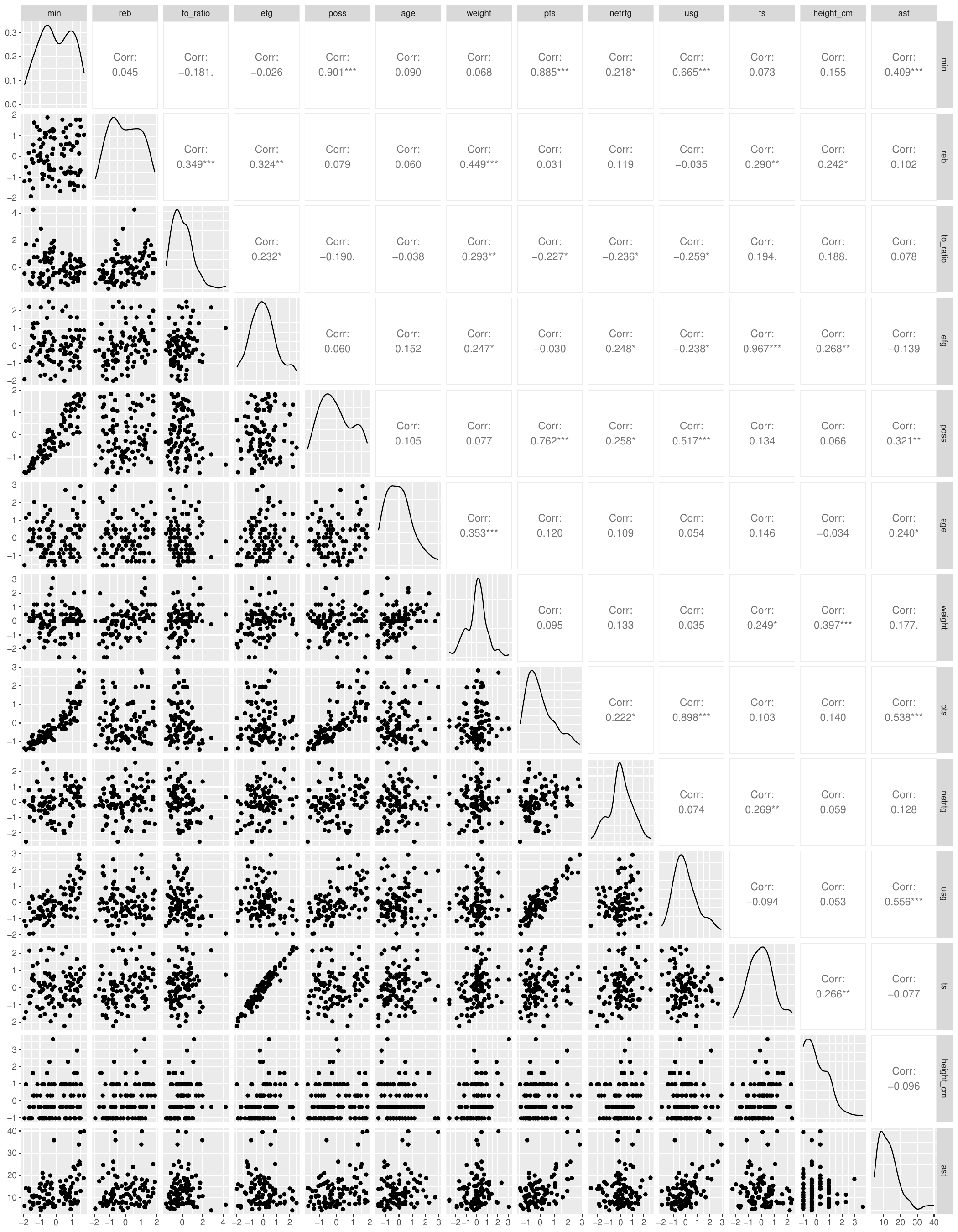} 
    \caption{Pairplot of reduced NBA dataset.}
    \label{fig:pairs}
\end{figure}

\begin{figure}
    \centering
    \includegraphics[scale = 0.3]{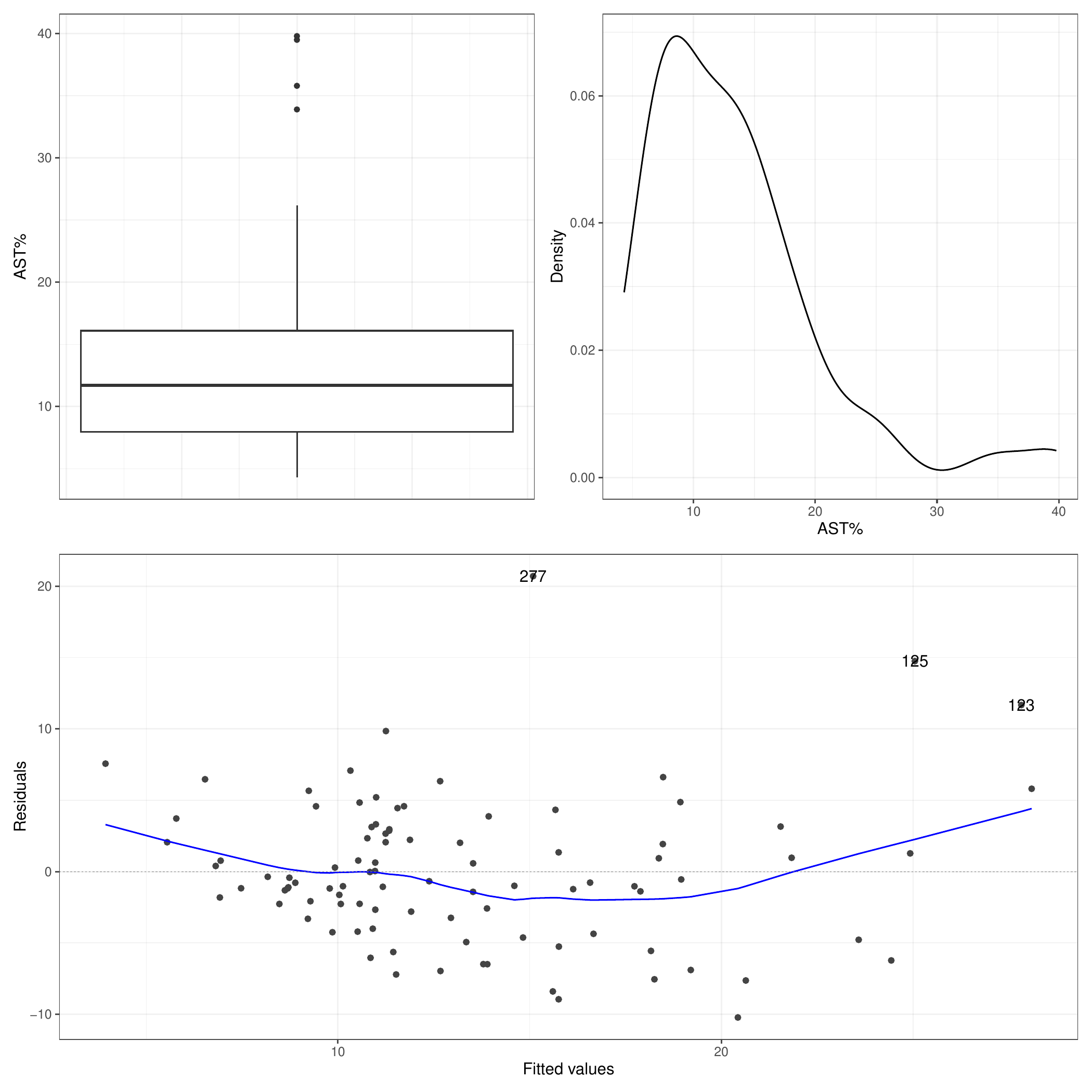} 
    \caption{Distribution and density of the NBA data response variable assist percentage.}
    \label{fig:NBAEDA}
\end{figure}

This appendix includes additional data information, including a table of metadata (Table \ref{tab:gloss}), a pairplot (Figure \ref{fig:pairs}) and a distribution analysis (Figure \ref{fig:NBAEDA}). The data was obtained from \url{https://www.nba.com/stats/players/bio?Season=2024-25&SeasonType=Regular+Season} and \url{https://www.nba.com/stats/players/advanced?Season=2024-25&SeasonType=Regular+Season}. The cleaned dataset can be found in the linked repository.

\begin{table}[h]
\centering
\renewcommand{\arraystretch}{1.2}
\begin{tabularx}{\textwidth}{lX}
\toprule
\textbf{Variable Name} & \textbf{Description} \\
\midrule
age & Age \\
ast & The percentage of teammate field goals a player assisted on while they are on the floor \\
efg & Efficient field goal - measures field goal percentage, adjusting for made 3-point field goals being 1.5 times more valuable than made 2-point field goals \\
gp & Number of games played \\
height & Height (cm) \\
min & Average minutes played \\
netrtg & Net rating - a team's point differential per 100 possessions while the player are on court \\
poss & Number of possessions\\
pts & Average points \\
reb & The percentage of a team's rebounds that a player has while on the court \\
to\_ratio & The number of turnovers a player averages per 100 possessions \\
ts & True shooting - a shooting percentage that factors in the value of three-point field goals and free throws in addition to conventional two-point field goals \\
usg & Usage - the percentage of team plays used by a player when they are on the floor \\
weight & Weight (lbs) \\
\bottomrule
\end{tabularx}
\caption{NBA metadata sourced from https://www.nba.com/stats/help/glossary.}
\label{tab:gloss}
\end{table}

\end{appendices}

\end{document}